\newcommand{\figpath}{./}

\newcommand{\ZZ}{{\mathbb Z}}               
\newcommand{\Frac}[2]{{\frac{\displaystyle\strut#1}{\displaystyle\strut#2}}}
\newcommand\lam{\lambda}
\newcommand\eps{{\varepsilon} }

\newcommand{\ineg}[2]{\nu_{#1,#2}}
\newcommand{\au}{AU}
\newcommand{\de}{ ^{\circ}}
\newcommand{\cA}{{\cal A}}
\newcommand{\cD}{{\cal D}}
\newcommand{\cDI}{{\cal D_{I^{\star}}}}
\newcommand{\cF}{{\cal F}}

\newcommand{\bsigg}{\bm{\sigma_g}}
\newcommand{\bsigs}{\bm{\sigma_s}}
\newcommand{\bun}{{\mathbf{1}}}
\newcommand{\pu}{{Paper\,I\,}}
\newcommand{\asay}{\text{ arcsec yr$^{-1}$}}

\def\be{\begin{equation}}
\def\ee{\end{equation}}

\newcommand{\FI}{{\it Family I}\;}
\newcommand{\FII}{{\it Family II}\;}
\newcommand{\FIII}{{\it Family III}\;}
\newcommand{\FIV}{{\it Family IV}\;}

\documentclass[usenatbib]{mn2e}

\usepackage[british]{babel}
\usepackage{graphicx,amssymb,amsmath,amsfonts,bm}

\begin{document}                                                                   

\title[Resonant structure of Jupiter's Trojans-II]{The resonant structure of Jupiter's trojan asteroids-II. What happens for different configurations of the planetary system. }

\author[P. Robutel and J.~Bodossian]
{P.~Robutel$^1$\thanks{E-mails: robutel@imcce.fr {\scriptsize (PR)}; 
bodossian@imcce.fr {\scriptsize (JB)}}
and J.~Bodossian$^1$\footnotemark[1] \\
$^1$ Astronomie et Syst\`emes Dynamiques, IMCCE, CNRS UMR 8028, Observatoire de Paris 77 Av. Denfert-Rochereau 75014 Paris, France\\
}


\maketitle

\label{firstpage}

\begin{abstract}
In a previous paper, we have found that the resonance structure of the present Jupiter Trojan swarms could be split up into four different families of resonances. Here, in a first step, we generalize these families in order to describe the resonances occurring in Trojan swarms embedded in a generic planetary system. The location of these families changes under a modification of the fundamental frequencies of the planets and we show how the resonant structure would evolve during a planetary migration. We present a general method, based on the knowledge of the fundamental frequencies of the planets and on those that can be reached by the Trojans, which makes it possible to predict and localize the main events arising in the swarms during migration. In particular, we show how the size and stability of the Trojan swarms are affected by the modification of the frequencies of the planets.   
Finally, we use this method to study the global dynamics of the Jovian Trojan swarms when Saturn migrates outwards. Besides the two resonances found by \citet{MoLeTsiGo2005} which could have led to the capture of the current population just after the crossing of the 2:1 orbital resonance, we also point out several sequences of chaotic events that can influence the Trojan population.
\end{abstract}


\section{Introduction}
\label{sec:intro}

The discovery of Achilles by Wolf in 1906, and of four other Jovian Trojans the next year, gave a new impulse to the study of the triangular configurations of the three-body problem, whose existence was shown by Lagrange in 1772. An important problem was to establish the existence of a stable area surrounding the triangular equilibrium points $L_4$ and $L_5$ associated to the Sun and Jupiter.  
From a mathematical point of view, the application of K.A.M. theory to the planar and circular restricted three-body problem, gives a result of confinement between two invariant tori, ensuring the stability in the neighborhood of $L_4$ and $L_5$ for an infinite time \citep{Leon1962,DeDe1967,Ma1972,MeSchmi1986}. 
In the case of the spatial restricted three-body problem, where K.A.M. theory does not ensure infinite time stability anymore,  \citet{BeFaGo1998} applied an extension of the Nekhoroshev theorem (\citeyear{Nekho1977}) to quasi-convex Hamiltonians in order to prove exponentially long-time stability.   But these two complementary theories do not give any information about the size of the stable region surrounding the triangular equilibrium points. In order to get estimates of this width, several authors developed Nekhoroshev-like estimates based on normalization up to an optimal order  \citep{GiDeFoGaSi1989,CeGio1991,GiSko1997,SkoDo2001}. More recent results, based on improvements of these methods, can be found in \citep{GaJoLo2005,EfSa05,LhoEftDvo2008}.
Another very interesting result was published by \citet{Ra1967}. In this work, based on the change of stability of the family of long-period Lyaponov orbits emanating from $L_4$, the author  gives an approximation of the limit eccentricity of stable tadpole orbits with respect to their amplitude of libration. 

Except Gabern and Jorba who applied the same methods as in \citep{GiDeFoGaSi1989}      to the bicircular, tricircular and biannular coherent problems \citep{Ga2003, GaJo2004},  and \citet{LhoEftDvo2008} who used a more sophisticated method in the elliptic restricted three-body problem (RTBP),  all of the above mentioned works were developed in the framework of the circular RTBP, planar or spatial, which is not a very realistic model for the purpose of studying the long-term dynamics of the Jovian Trojans. To overcome this problem, several authors performed full numerical integrations.  In \citeyear{HoWi1993}, \citeauthor{HoWi1993}, integrating test-particles during 20 Myr, obtained the first global stability result concerning the Trojans of the giant planets in our Solar system.  The same year, \citeauthor{Milani93} developed a numerical method providing the proper elements of the Trojans of Jupiter. A  semianalytical  model of determination of these proper elements has also been proposed by \citeauthor{BeaugeR01} in  \citeyear{BeaugeR01}.
Four years later, \citet{LevisonSS97} described the long-time erosion of the Jovian Trojan swarms without identifying its mechanism (more results concerning the long-term behavior of the Trojans are published in   \cite{TsiVaDvo2005}).
A   few years later, \citet{MiBeRo2001},  \citet{NeDo02} and \citet{MaTriSch2003}, showed the existence of two sets of unstable structures lying inside the Jovian Trojan swarms which could be related to the  long-time erosion. The authors suggested that one of these sets was connected to the great inequality (between Jupiter and Saturn) while the second one may have been generated by the commensurability between the libration frequency of the co-orbital resonance and the frequency $n_{Jup} - 2n_{Sat}$.    
 In \citeyear{RoGaJo2005},  \citeauthor{RoGaJo2005} identified these singularities and analyzed their underlying resonances, this leading to the decomposition of the "resonance structure" in four families of resonances.   In \citep{RoGa2006}, hereafter called \pu, a new description of this resonant structure was given, and the link between these resonances and the long-term stability (or instability) of Jovian Trojans was established.  
 
The structures described in \pu, generated by commensurabilities between the proper frequencies of the Trojans and the fundamental frequencies of the planetary system, depend on the planetary configuration. Therefore, a small modification of the geometry of the system is able to modify the Trojan swarms' resonant structure, changing the global dynamics, and, consequently, the stability of the co-orbital region.  
Planetary migration in a planetesimal disc (see  \citet{GoMoLe2004} and references therein) provides a natural mechanism of evolution of planetary systems. 
According to \citet{MoLeTsiGo2005}, the Jovian co-orbital population is not primordial, but was, instead, captured during chaotic events which took place in the course of a planetary migration.   
The so-called Nice model \citep{TsiGoMoLe2005} predicts that during the migration (inwards for Jupiter, and outwards for Saturn, Uranus and Neptune), the  Jupiter-Saturn couple crossed the 2:1 orbital resonance. \citet{MoLeTsiGo2005} suggested that the Trojans were captured in co-orbital motion just after the crossing  of this resonance, when the dynamics of the Trojan region was completely chaotic.
The authors show that this global chaos arises when the libration frequency of the Trojans is close to the combinations  $3(n_{Jup} - 2n_{Sat})$ or  $2(n_{Jup} - 2n_{Sat})$.
This result was recently confirmed by  \citet{MaScho2007}, showing that these resonances are the main generators of the depletion (and, probably also the capture) of the Trojan population. 

The results reported in \pu are directly related to the phenomenon described in  \cite{MoLeTsiGo2005} and  \citet{MaScho2007}. More precisely, the resonances involved in this process are members of one of the four families of resonances presented in \pu. Consequently, this previous paper contains the necessary material to derive tools, that make it possible to study  the evolution of the Trojans resonant structure. This evolution is induced by changing the geometry in the considered planetary system. Indeed, our goal is not to give an accurate and realistic description of the behavior of the Trojan swarms during a migration process, but rather to study what are the planetary configurations for which the Trojan swarms become globally chaotic. Moreover, we want to develop a model that is capable of making predictions in a large class of problems, and that is not restricted to the study of the Jovian Trojans. 
Our approach is based on three different points.
The first one rests on the understanding of the Trojans' resonant structure, and its decomposition. 
It will be shown in section \ref{sec:generical} that, for a generic Trojan swarm, almost all resonances driving its global dynamics are members of four different families which generalize the families described in \pu.
The resonances of all these families involve both fundamental frequencies of the Trojans and the basic planetary frequencies. These families establish a link between the behavior of the Trojan swarm and the geometry of its planetary system. 
The two other ingredients  on which the method is based are respectively the exploration of the planetary frequencies and the determination of the frequency domain to which  the main frequencies of the whole Trojan swarm belong. 
In section \ref{sec:Jupiter}, our method is applied to the study of the Jovian Trojans perturbed by Saturn, employing the model  used in \pu but also in \citep{MoLeTsiGo2005,MaScho2007}. 
 This application reveals the rich dynamics of the Trojan swarms under the action of numerous resonances of different origins, which generate a resonance web which cannot be fully understood without knowing the fundamental frequencies of the considered dynamical system.

\section{The resonance structure of a general Trojan swarm}
\label{sec:generical}
\subsection{ General setting }
\label{sec:general}
Let us consider a general case:    the planet harboring  the Trojan swarms  is the $p^{th}$ planet of a given planetary system composed of $N$ planets orbiting a central star. 
We denote by $m_j$, respectively  $a_j$, the masses and the semi-major axes of the planets, and $m_0$ the mass of the star.  The small mass parameter $\mu$ is defined by the expression  $\mu = \text{max}(m_1/M, \cdots, m_N/M)$, where M  is the total mass of the system. A linear combination of secular frequencies will be denoted:   ${\bf k_g}\cdot \bsigg +  {\bf l_s}\cdot \bsigs$ where ${\bf k_g}$ and ${\bf l_s}$ are elements of $\ZZ^N$.
If we assume that the planetary system is stable enough to be considered as quasiperiodic for a given time length (see paper I), its fundamental frequencies are denoted:
$(n_1, \cdots, n_N)$, $\bsigg =  (g_1, \cdots, g_N)$ and $\bsigs = (s_1, \cdots, s_N)$ \footnote{ Owing the invariance of the total angular momentum, one of the $s_j$ is equal to zero.}, where the $n_j$ denote the proper mean motions  which are of order $0$ with respect to the planetary masses. The vectors $\bsigg$ and  $\bsigs$ correspond to the secular frequencies of the planetary system (respectively associated to the precession of the perihelia and to the precession of the ascending nodes) which are both of order  $\mu$. 

\subsection{ Restricted three-body problem }
\label{sec:RTBP}

In a first step, we retain only the gravitational interaction of the $p$'th planet (and of the star) on the Trojans. We are thus brought back to the RTBP. This model gives, at least for the outer planets of the Solar system, results which are quite realistic (this is not the case for  inner planets, see section \ref{sec:RNp2BP}). In this model, the triangular equilibrium points  $L_4$ or $L_5$  are well defined, even for an elliptic motion of the planet (paper I). Strictly speaking, in an inertial frame, the equilateral configurations correspond to periodic orbits. But in a suitable reference frame, which is an uniformly rotating frame in the circular problem, and a non uniformly rotating and pulsating frame in the elliptic problem (see \cite{sze1967}),  these periodic orbits become fixed points.  When the motion of the secondary is circular, these fixed points are elliptic (linearly stable) as long as the inequality $27m_pm_0<(m_p+m_0)^2$ is satisfied \citep{Ga1843}. This ensures the linear stability of the equilateral equilibria as soon as $\mu$ belongs to the interval $[0,\mu_0]$, with  $\mu_0 = (1-\sqrt{23/27})/2 \approx 0.0385$.     For eccentric motions of the secondary, the (linear) stability criterion depends on its eccentricity, and stability exists for values of $\mu$ greater than $\mu_0$ (see \citep{Dan1964,Robe2002} for numerical estimates and  \citep{MeSchmi2005} for analytical ones) . 
When the equilibrium is linearly stable, the eigenfrequencies (moduli of the eigenvalues)  $ \nu_{L_{4,5}} $, $g_{L_{4,5}}$  and $s_{L_{4,5}}$, which yield the fundamental frequencies associated to libration, precession of the perihelion and precession of the ascending node of the asteroid, are given by: 
\be
\begin{split}
 & \nu_{L_{4,5}}  = \sqrt{(27/4)\eps}n_p +o(\sqrt{\eps}) = O(\sqrt{\mu}),  \\
 & g_{L_{4,5}}  = (27/8)\eps n_p +o(\eps) = O(\mu) \\
 & s_{L_{4,5}}  = 0  \quad \text{ with}\, \eps = \Frac{m_p}{m_0 + m_p} =   O(\mu). 
\end{split}
\label{eq:freq_L4}
\ee

General theory gives the values of the frequencies not only at the Lagrangian points, but also in the tadpole region as well as for horseshoe orbits \citep{Gar1976a,Morais2001}. According to these theories, the magnitude of the frequencies remains unaltered in the whole phase  space, except for $s$  which reaches the order $O(\mu)$ but remains always lower than $g$ (in absolute value). 
Owing to these three different time scales, significant resonances between these frequencies are very unlikely (at least asymptotically). In the present Solar system, these resonances are negligible as long as $\mu$ is lower than $10^{-3}$ (close to Jupiter mass).
Several regions of Jupiter's Trojan swarms are deeply affected  by secondary resonances involving the commensurability of the libration frequency with Jupiter's proper mean motion. In \pu, these resonances, elements of \FI, are defined by:

\begin{equation}
i\nu + j n_p + k g= 0, \qquad (i,j,k)\in \ZZ^3.
\label{eq:sec_ress_ertbp}
\end{equation}

For Jupiter's Trojan swarms, among this family, the influence of the resonances satisfying $j=1$ and $i \in\{12, 13, 14\}$ is dominant for moderate to high libration amplitude.
In the present paper, these three resonances can be seen in Fig.  \ref{fig:frec_FIV_I02} (see section \ref{sec:RNp2BP}).  Although the dynamical implication of these resonances is weaker in the circular RTBP than in the elliptic one, \FI is well known in the first model. Indeed, the existence of denominators associated to the resonances  $(i,j) \in \{(11:-1), (12:-1), (13:-1), (14:-1)\}$ during the Birkhoff normalization process is already mentioned in  \citep{DepritHR67} . But these terms do not generate any difficulty up to degree $15$.  Moreover, those resonances are identified in \citep{GiDeFoGaSi1989} as the factor of divergence of the normal forms in the spatial circular RTBP.   \citet{ErNaSaSuFro2007} study in detail these secondary  resonances in the  elliptic RTBP and their dependence on the mass and eccentricity of the secondary.

\subsection{ Restricted (N+2)-body problem }
\label{sec:RNp2BP}

The first difficulty comes from the fact that, in the restricted (N+2)-body problem, since $N$ is greater or equal to $2$, the triangular equilibrium points do not exist. These points are replaced by elliptic quasiperiodic trajectories (see \citep{JoSi1996,Jo2000,Ga2003}).   But numerical experiments show that, if the additional planetary perturbations are not too large, a stable region remains in the neighborhood of the equilateral points. As in the RTBP, these equilateral points, which are always properly defined, are usually called $L_4$ and $L_5$.

This point being clarified, let us study the influence of planetary perturbations on the Trojans associated to the $p^{th}$ planet.  
These dynamical effects on the Trojan swarms can be mainly split in two classes: the direct influences, due to the gravitational attraction of the planets on the Trojans; and the indirect effects, coming from the fact that the $p^{th}$ planet does not evolve any more on a Keplerian trajectory\footnote{In order to simplify the following discussion, we approximate all the motions as quasiperiodic, which is not necessarily the case.}. Although this difference between direct and indirect perturbation is quite arbitrary (we will see later that in some cases these two phenomena are mixed together), it has the advantage to simplify the discussion.    
In terms of frequencies, the main effect of the direct influences is to modify the proper frequencies $(\nu,g,s)$ of the Trojans. Assuming that the motion of the additional planets is circular (we consider here the $N$-circular problem), according to the Laplace-Lagrange theory, the secular linear contribution of the $j^{th}$ planet to the proper precession frequency $g$ can be roughly approximated at $L_4$ (or $L_5$) by the expression:
\begin{align}
g^{(j)}_{L_{4,5}} =  { \frac14 }n_p\mu_j\alpha_jb_{3/2}^{(1)}(\alpha_j), \quad \text{ if }\, \alpha_j={\frac{a_j}{a_p}}<1.\\
g^{(j)}_{L_{4,5}} = { \frac14 }n_p\mu_j\alpha_j^2b_{3/2}^{(1)}(\alpha_j), \quad \text{ if }\, \alpha_j={\frac{a_p}{a_j}}<1. 
\label{eq:pert_plan}
\end{align}
where $b_{3/2}^{(1)}$ is a Laplace coefficient and $\mu_j = m_j/m_0$ (see for example \citep{MuDe1999}).
Then, the secular frequencies at the triangular equilibrium points are given by: 

\begin{align}
g^{(Tot)}_{L_{4,5}}  =  g^{(R.T.B)}_{L_{4,5}} + \sum_{j\neq p} g^{(j)}_{L_{4,5}} \\
s^{(Tot)}_{L_{4,5}}  =  s^{(R.T.B)}_{L_{4,5}} - \sum_{j\neq p} g^{(j)}_{L_{4,5}} 
\label{eq:pert_plan_tot}
\end{align}
where $g^{(R.T.B)}_{L_{4,5}} = (27/8)\eps n_p$ and  $s^{(R.T.B)}_{L_{4,5}} =0$, see formulas (\ref{eq:freq_L4}).
The frequencies  $g^{(Tot)}_{L_{4,5}}$ and $s^{(Tot)}_{L_{4,5}}$ as well as $ g^{(R.T.B)}_{L_{4,5}}$ and the planetary contribution $g_{L_{4,5}}^{(plan)} = \sum_{j\neq p} g^{(j)}_{L_{4,5}}$ are gathered in Table (\ref{tab:freqL4}), for every planet of the Solar system.  Although the results of Table (\ref{tab:freqL4}) are valid only at the triangular points and for circular and coplanar planetary orbits, they are sufficient to deduce striking conclusions.  
As we can see by comparison of the second and third column, the planetary contribution $g_{L_{4,5}}^{(plan)}$ is not always a perturbation, but, at least for Mercury, Mars and Uranus, it can be the main contribution to the precession frequencies. It is only for Saturn and especially for Jupiter that the direct perturbations by the other planets impose only small corrections to their Trojans' secular frequencies.  
Consequently, except for Jupiter, the planetary direct gravitational attraction drastically modifies the dynamics predicted by the RTBP.
This phenomenon has been emphasized by  \citet{EvTa2000a} in the case of Mercury's Trojans. Indeed, in the full Solar system model, the authors found that the most stable zones (100 Mys stability) did not contain the Lagrange points. 
Another striking point is that, in an outer Solar system simulation, the region of smallest amplitude of libration of Uranian and Neptunian Trojans is shifted by more than $0.1 \au$ from its predicted location in the RTBP \citep{NeDo02}.

The indirect perturbations act in a more subtle way than the direct ones: the introduction of forcing frequencies  (i.e. the planetary frequencies, which are constant once the planetary system is given) enables resonances between the Trojan's frequencies and these additional frequencies.
Obviously, the secondary resonances with the orbital frequency $n_p$ are directly affected by the additional secular frequencies. These forcing frequencies increase the number of possible resonant harmonics, imposing a generalization of \FI by: 
\be
i\nu + jn_p = -( kg + ls + {\bf k_g}\cdot \bsigg +  {\bf l_s}\cdot \bsigs)\, .  \quad
 \label{eq:FI}
\ee
with $i\neq 0$, $j\neq 0$ and $j + k + l  + {\bf k_g}\cdot \bun + {\bf l_s}\cdot \bun = 0$.
In these expressions the dot denotes the Euclidian scalar product and $\bun = (1, \cdots, 1)$.
The enrichment of this family may generate large chaotic zones due to the overlapping of those resonances. 
Another expected consequence is the introduction of secular resonances. In the Solar system, as is shown by the last two columns of Table \ref{tab:freqL4}, except for the frequency $g$ of Jupiter's and Saturn's Trojans,  the secular frequencies at $L_4$ are very close to the fundamental frequencies of the perihelia and of the nodes of the planets \citep{La90,LaRoJoGaCoLe2004}.  Indeed, the resonance $s=s_2$ which limits the long-term stability region  of low inclination Jovian Trojans\footnote{Here, $s_2$ has to be understood as the fundamental frequency associated to the precession of the ascending node of Saturn. This frequency is usually denoted $s_6$.} was first mentioned by \citet{Yoder79}, other secular resonances involving $s$ are discussed in \citep{Milani93,Milani94}. The role of secular resonances in the motion of the Trojans of the inner planets was investigated by \citet{BraLe2002}.  More accurate studies were dedicated to the role of secular resonances in the motion of Venus Trojans \citep{Michel1997,SchoMaTri2005} and Mars Trojans \citep{SchMaTri2005}.  As in \pu, the family containing secular resonances is denoted \FIII and defined by:
\be
k g + l s + {\bf k_g}\bsigg + {\bf l_s}\bsigs = 0.
 \label{eq:FIII}
\ee
Contrarily to \pu, where $k$ is always set equal to zero,  the possibility that $ k\neq 0$ in Jupiter's Trojan swarms is discussed in sections \ref{sec:plan-freq} and \ref{sec:FII}.  

The conjunction of both direct perturbations and secular resonances prevents inner planets, except Mars \citep{SchMaTri2005}, from having long-lived Trojans. For this reason, the Trojans of the first three planets of the Solar system are generally transient objects, which spend only a few hundreds of thousands years in the co-orbital region \citep{MoMo2002, MoMo2006}.

\begin{table}
\caption[]{ Linear secular frequencies evaluated at the triangular points of the planets of the Solar system, expressed in $\asay$. In the first column is the name of the planets. The second one gives the precession frequency of the perihelion derived from the circular RTBP (formula (\ref{eq:freq_L4})), while the third one gathers the contribution of the planets to this frequency (formula (\ref{eq:pert_plan_tot})).  Finally, the precession frequencies $g$ and $s$ at $L_4$ or $L_5$ are given in fourth and fifth columns.}
\begin{tabular}{r r r r r }
\hline
\noalign{\smallskip}
   &  $g_{L_{4,5}}^{(R.T.B.)}$ & $g_{L_{4,5}}^{(plan)}$  &  $g_{L_{4,5}}^{(Tot)}$ & $s_{L_{4,5}}^{(Tot)}$ \\
  Me & $  3.0$   &$  5.5$ &$8.5$ & $-5.5$\\  
   V   & $ 17.3$  & $ 12.2$ & $29.5$ & $-12.2$ \\  
   E   & $ 13.3$  &$12.9$ &$26.2$ &$-12.9$ \\  
   Ma & $  0.7$  &$17.8$ &$18.6$ & $-17.8$\\  
   J    & $352.3$ &$7.4$ &$359.8$ &$-7.4$ \\  
   S   &$ 42.3$   &$18.3$ &$60.6$ &$-18.3$ \\  
    U  & $  2.3$   & $2.7$& $5.0$&$-2.7$ \\  
    N  & $  1.4$   &$0.7$ &$2.0$ & $-0.7$\\  
\end{tabular}
\label{tab:freqL4}
\end{table}

Until now, only secular forcing frequencies have been taken into consideration, but combinations of planetary mean-motions also play a major role in some specific configurations of the planetary system. If we consider a $\beta:\alpha$ MMR between the $p^{th}$ planet and another one\footnote{MMR involving three and more bodies  are not taken into account here.}, let us say the $q^{th}$, its critical angle reads:
\be
\theta = \alpha\lam_p - \beta\lam_q + \cdots  ,
\ee
 where the dots represent  a linear combination of longitudes of the nodes and of the perihelia such that the d'Alembert rules are satisfied. The modulus of the quantity $\ineg{\alpha}{\beta} = {\alpha}n_p - {\beta}n_q$ will depend on how close the planetary system is to the resonance. The closest  the system will be to the resonance, the smallest $\vert \ineg{\alpha}{\beta}\vert $ will be. 
Hence, far from the MMR, $\ineg{\alpha}{\beta} = O(1)$ (same order as the planetary mean motions). 
The quantity $\vert\ineg{\alpha}{\beta}\vert$ becoming smaller and smaller as the two planets approach the exact resonance, this frequency will reach values that can generate resonances with the fundamental frequencies of the Trojans. As the planetary system approaches the MMR,  $\vert\ineg{\alpha}{\beta}\vert$  will at first be of the same order of magnitude as $\nu$ (i.e. $0(\sqrt{\eps})$), enabling commensurabilities which generate the resonances of  \FII, defined by:
\be
   i\nu  -  j\ineg{\alpha}{\beta} = -( kg + ls + {\bf k_g}\cdot \bsigg + {\bf l_s}\cdot \bsigs) 
 \label{eq:FIIp}
 \ee 
 with $j(\beta - \alpha) +k + l + {\bf k_g}\cdot \bun + {\bf l_s}\cdot \bun = 0$.
This generalizes \FII as it is defined in \pu.
Once this threshold is crossed, no new significant resonance arises until the planets are very close to the MMR. Here, $\ineg{\alpha}{\beta}$ is of the same order as $g$ (i.e. $O(\eps)$), which generates the resonances of \FIV satisfying:
 \be
 j\ineg{\alpha}{\beta} +   kg  = -(  ls + {\bf k_g}\cdot \bsigg + {\bf l_s}\cdot \bsigs )
 \label{eq:FIV}
 \ee
with $j(\alpha - \beta) +k + l + {\bf k_g}\cdot \bun + {\bf l_s}\cdot \bun =0$. This generalizes the family \FIV  that is defined in \pu.

A similar phenomenon arises after the crossing of the orbital resonance, when the frequency $\vert\ineg{\alpha}{\beta}\vert$ increases from zero to $O(1)$.

\section{Application to Jupiter's Trojans}
\label{sec:Jupiter}
\subsection{One parameter model}
\subsubsection{Model and method of analysis}
\label{sec:model}

\begin{figure}
\includegraphics[width=8.3cm,angle=0]{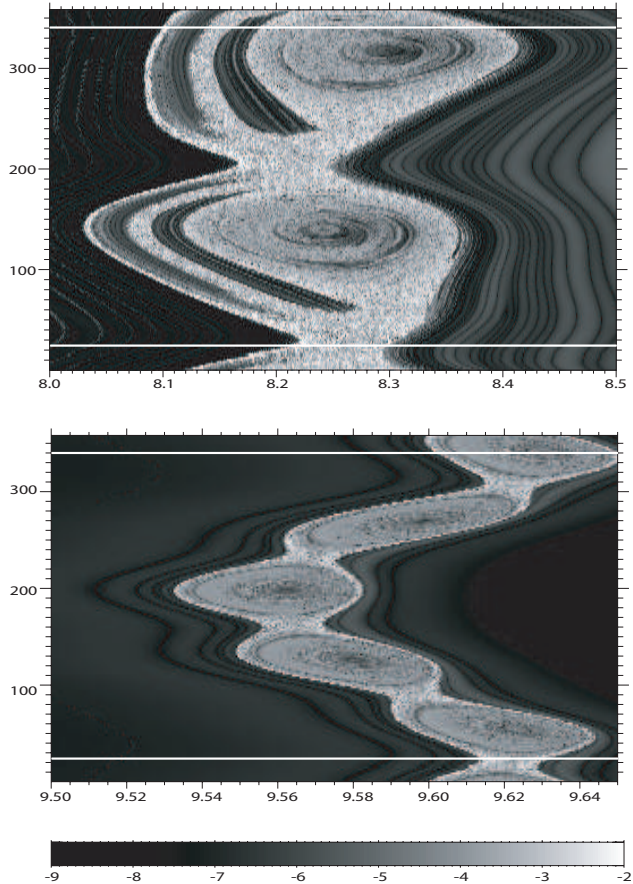}\\
\caption[]{ Section of the phase space of the planetary system by the plane of initial conditions $(a_2,M_2)$. Neighborhood of the 2:1 (top) and 5:2 (bottom) mean motion resonances. The small bottom strip displays the grey code associated to the diffusion index: the darker the grey, the more regular the trajectories. The two horizontal white lines correspond respectively to the "elliptic segment" for $M_2=340.04^{\circ}$ and to the "hyperbolic segment"  for $M_2=24.14^{\circ}$ }
\label{fig:res_plan}
\end{figure}

In this section, the evolution of the resonant structure in the Trojan swarms mentioned in section \ref{sec:RNp2BP} will be illustrated using a concrete planetary system. To this purpose, the methods described in \pu will be followed. 
We consider the planetary system made of the Sun, Jupiter and Saturn. In order to be consistent with the previous section, the index $1$ is associated to Jupiter and $2$ to Saturn.
In \pu, the system, in its present configuration, is close to the 5:2 MMR. Consequently, resonances of \FIV should be present in the Trojans' phase space.   Indeed, several resonances of this family (for $j=1$ and $k=4$ in formula  (\ref{eq:FIV})), have been clearly identified in \pu.
 We have shown in this paper, that these resonances are associated to their long-term erosion of the Trojan swarm.

Despite the long distance between the theoretical location of the 2:1 resonance and the present couple of giant planets (more than $1.2\au$) resonance members of \FII associated to $\nu_{2,1}$ have also been clearly identified.  Owing to this distance, the combinations between $\nu$ and $\ineg12$ are of high order: $i=5$ and $j=-2$ in formula (\ref{eq:FIIp}).
Obviously, secondary resonances of \FI and secular resonances of \FIII play a major role in Jupiter's present swarms of Trojans (\pu). But, as these families are only weakly affected by migration (section \ref{sec:plan-freq}), except when they are very close  to a MMR (section \ref{sec:FII}), we will not really pay attention to these resonances.

In order to appreciate the modifications of the resonant structure of Jupiter's Trojans due to different relative positions (and distances) between the two planets, we consider a sequence of independent planetary initial conditions. The fundamental frequencies of the planetary system depending mainly on the semi-major axes of the planets, or more precisely on their ratio, we have decided to change only one parameter: the semi-major axis of Saturn $a_2$. Hence, we are left with a one-parameter model. 
More concretely, for every value of the parameter $a_2$, we integrate the Sun, Jupiter, Saturn and a set of fictitious Jupiter's Trojans (considered as test-particles). 
Except for the initial semi-major axis of  Saturn, which is chosen between $8$ and $9.7$ \au, the initial conditions of Jupiter and Saturn are the ones given by DE405 at the Julian date 245 2200.5 (2001 October 10). Regarding the Trojan swarms, the initial elements of their members are independent of the parameter. In each run, a grid of initial conditions is considered with $200$ initial values of the semi-major axis and $40$ values of the eccentricity equally spaced in the domain $\cA=[5.2035,5.4030]\times[0.05,0.30]$ ($8000$ test-particles), while the other elements are fixed to the following values: $\sigma = \lambda - \lambda_1 = \pi/3$, $\sigma_g = \varpi - \varpi_1 = \pi/3$, $\Omega = \Omega_1$ and $I = I_1+I^{\star}$. Because the resonant structure depends on the initial inclination (\pu), $I^{\star}$ is fixed to three different values: $2^{\circ}$, $20^{\circ}$ and $30^{\circ}$. In this paper, this set of initial conditions is denoted $\cDI$.

The numerical simulations are performed by the symplectic integrator $SABA_4$ \citep{LaRo2001} with an integration step of 1/2 year. Trojans and planets are integrated on two consecutive time-spans of $5 Myr$. For particles surviving the integration (bodies which are not ejected from the co-orbital region before the end of the integration) the fundamental frequencies are computed for each of these two time-intervals using the frequency analysis method developed by \citet{La90}. If we denote by $\cF$ the map which  associates to each Trojan of $\cDI$ its fundamental frequencies $(\nu,g,s)$ (see \citealt{La99} and \pu), the domain of the frequency space reached by the Trojans is: $\Theta_{I^{\star}} = \cF(\cDI)$. 
Consequently, three types of complementary information are derived from the study of $\cDI$.  in \pu
The most straightforward piece of information is given by the escape rate: the number of Trojans escaping the co-orbital region before the end of the integration ($10$ My), divided by the initial number of Trojans inside  $\cDI$.  This indicator will be widely used in section \ref{sec:FII}. Owing to a loss of accuracy of our integrator during close encounters with Jupiter, the depletion of   $\cDI$ is probably over-estimated, but in any case, this ejection rate is always correlated to the global instability of the considered region.
    Other significant information can be found in the frequencies. Since a detailed discussion of the application of Frequency Map Analysis to the Trojans can be found in \pu, let us mention two applications of this method. First, the comparison of the set of frequencies computed on the two intervals of $5 Myr$ makes it possible to derive the diffusion rate of every Trojan (index related to its stability). Practically, if we denote by $\nu^{(1)}$ the libration frequency determined on the first time span, and $\nu^{(2)}$ the same quantity computed on the second interval, the diffusion index will be given by the quantity: $\log_{10}\vert\Frac{\nu^{(1)}- \nu^{(2)}}{\nu^{(1)}}\vert$. From this index, a dynamical map of the domain $\cDI$ is derived. Second, the study of the frequency domain $\Theta_{I^{\star}}$ enables us to understand how the chaotic regions are generated by the overlapping of the underlying resonances.  These complementary techniques will be used in section \ref{sec:crossingFIV52}.

Once the fundamental planetary frequencies are known,  it is straightforward to predict whether the families of resonances associated with these frequencies are inside the Trojans' phase space or not.  Obviously, in order to make these predictions, the bounds of the frequency domain $\Theta_{I^{\star}}$ have to be known. To this aim, we assume that these bounds do not depend on the value of $a_2$, which is a very good approximation in the case of Jupiter's Trojans. As it was established in \pu, we assume in the following sections that:
 \be
 \begin{split}
  &\Theta_{2\phantom{0}}     \subset  [7700,9150] \times [310,445] \times[-45,-7.5] \\
  &\Theta_{20} \subset  [7400,8660] \times [285,350] \times[-40,-3.5]\\
  &\Theta_{30} \subset  [7000,8138]\times [251,280]     \times[-30,+0.6]
 \end{split}
\label{eq:bounds}
\ee
where the three intervals are respectively the projections of $\Theta_{I^{\star}}$ on the 1-dimensional space of $\nu$, $g$ and $s$ denoted $\pi_{\nu}(\Theta_{I^{\star}})$, $\pi_{g}(\Theta_{I^{\star}})$ and $\pi_{s}(\Theta_{I^{\star}})$. The units of frequency are $\asay$. It is important to mention that the lower bounds of $\nu$ and $s$ here are arbitrary. For example, at $2^{\circ}$ of initial inclination, the Trojans whose libration frequency is lower than about  $7800\asay$ have trajectories relatively far from quasi-periodic, making the use of fundamental frequencies less significant (details can be found in \pu).  The same remark holds when $s$ is lower than the bounds indicated in (\ref{eq:bounds}). 
These bounds may also be compared to those obtained by analytical fits of $(\nu,g,s)$ \citep{Milani94,MaTriSch2003}, which give quite similar results up to $30^{\circ}$ of initial inclination.

\subsubsection{ Behavior of the planetary frequencies}
 \label{sec:plan-freq}

\begin{figure*}
\includegraphics[width=11cm,height=16cm,angle=270]{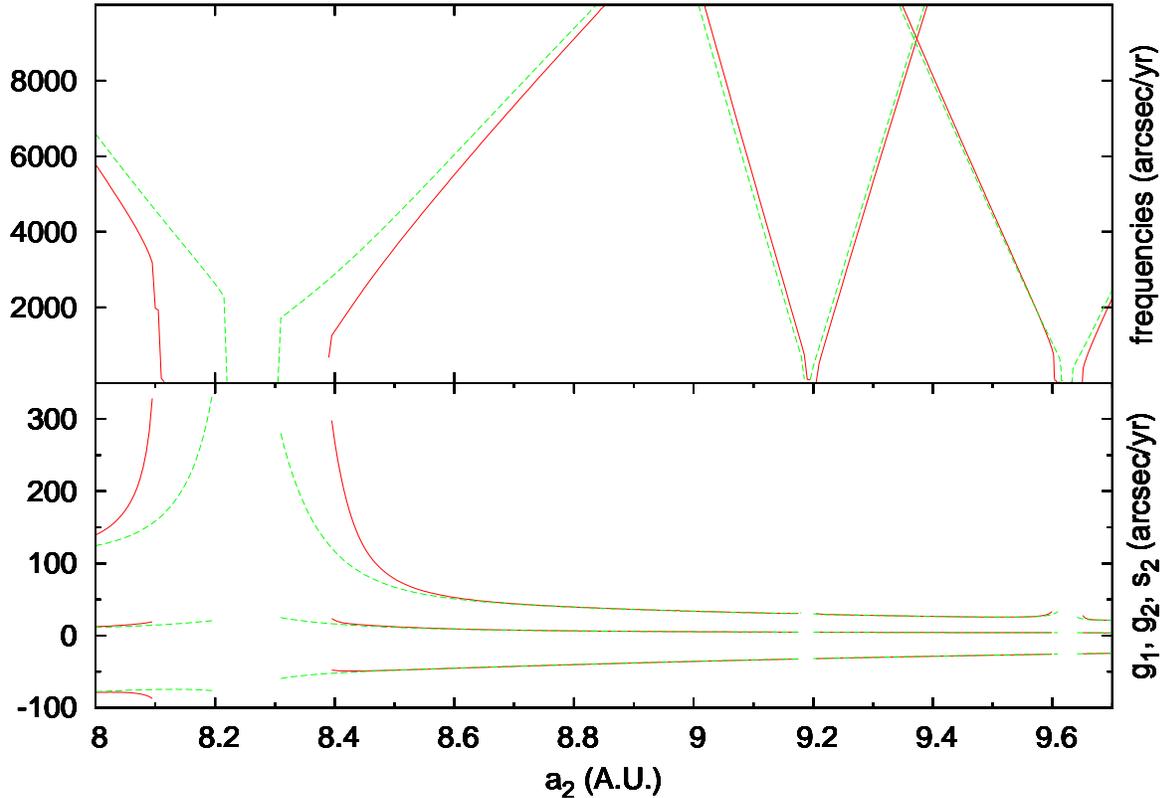}
\caption[]{Fundamental frequencies of the planetary system versus $a_2$. Top: combinations of frequencies $\vert \ineg12\vert$,  $\vert \ineg37\vert$ and  $\vert \ineg25\vert$ which are associated to the resonances of \FII.      Bottom: secular frequencies $g_1, g_2$ and $s_2$. The empty intervals correspond to the 2:1, 7:3 and 5:2 MMRs where the secular frequencies are singular.  The frequencies associated to the "elliptic" section of the phase space ($M_2 = 340.04\de$) are plotted in red, while the "hyperbolic" section  ($M_2 = 24.14\de$) is in green. See the text for more details.}
\label{fig:frec_plan}
\end{figure*}

From now on, Jupiter's and Saturn's semi-major axes and  proper mean motions  are denoted $(a_1,a_2)$ and $(n_1,n_2)$. Following the notations established in section \ref{sec:general}, the secular frequencies of the planetary system are $\bsigg =(g_1,g_2)$ and $\bsigs = (0,s_2)$.  These frequencies play a fundamental role in this study. It is therefore essential to know their variations with respect to the parameter $a_2$.
We have seen, in section \ref{sec:RNp2BP} that the location of the planetary system with respect to MMRs plays a major role in the transition from the resonances of \FII to the ones of  \FIV.  
It is therefore important to know the geometry of the orbital resonances in the phase space.
 Fig. \ref{fig:res_plan} shows dynamical maps of the regions surrounding the orbital resonance 2:1 (top) and 5:2 (bottom), corresponding to the section of the phase space by the plane of coordinates $(a_2,M_2)$, $M_2$ being the mean anomaly of Saturn.   The grey code indicates the diffusion rate of Saturn's proper mean motion computed on two consecutive time-intervals (section \ref{sec:model}). From these two determinations, denoted respectively $n_2^{(1)}$ and $n_2^{(2)}$, a diffusion index is derived by the expression: $\log_{10}\vert ( n_2^{(1)} -  n_2^{(2)} ) /n_2^{(1)} \vert $.
  The structures of high diffusion rate (light grey for diffusion rate greater than $-3$) are associated to the inner part of the resonance (libration island). Fig. \ref{fig:res_plan} shows the characteristic shape of the resonant chains composed of two islands for the 2:1 and five for the 5:2 (see \citet{RoLa2001} for details). These large islands, which correspond to globally stable regions (elliptic regions in the pendulum model), are separated by narrow unstable structures like the hyperbolic fixed points in the simple case of the pendulum. Even if the description of the dynamics of this problem is outside the scope of this paper, two points are interesting to note. First, the dynamics inside the resonant islands appears very rich. Indeed, very sharp structures indicated by different diffusion rates are clearly visible; they are probably related to secondary or secular resonances.  Second, particularly for the 5:2 MMR, the island chain is strongly distorted, making difficult to define a resonance width.
The segment containing the initial conditions used for our simulation is represented in Fig.  \ref{fig:res_plan}.   Along this segment, the initial values of $M_2$ are always equal to  $M_2 = 340.04^{\circ}$. This line of initial conditions crosses the first lobe of the 2:1 orbital resonance nearly along its widest section, and passes near the libration center.  For this reason, we call it "elliptic segment".
According to \citet{MoLeTsiGo2005}, during planetary migration, the MMRs, and particularly the 2:1 are "jumped" by the planetary system. More precisely, if the migration is slow enough to satisfy the adiabatic invariance hypothesis, the system has to cross the resonance through its hyperbolic fixed point (this makes sense for one degree of freedom systems) without reaching any libration zone (the same phenomenon is observed in \citealt{MaScho2007}).
In order to compare the "elliptic crossing" to the "hyperbolic" one, we consider a second set of initial conditions represented in Fig.  \ref{fig:res_plan} by the second horizontal white line ($M_2 = 24.14^{\circ}$). This path will be called "hyperbolic segment". We also notice in Fig.\ref{fig:res_plan} (bottom)  that this segment passes through the hyperbolic region connecting the lobes of the 5:2 resonance.  Keeping  this figure in mind, it is now easy to understand the behavior of the planetary frequencies under variations of $a_2$ along the two above mentioned paths. 

Jupiter's proper mean motion $n_1$ is practically not affected by the variations of $a_2$. 
 As a result, the location of the resonances of \FI is almost independent of the value of $a_2$. Only small shifts are observable due to the variation of the secular frequencies involved in formula (\ref{eq:FI}). 
  On the contrary, $n_2$ being affected by the variation of the parameter, the mean motion combinations associated to orbital resonances (here $\ineg12$,  $ \ineg37$ and  $ \ineg25$) vary drastically\footnote{ Only these three resonances are taken into account in our study. Other MMRs of higher order like the 9:4, 11:5 and 12:5 MMRs are also present in the domain of study, but their influence on the Trojans remains too small to be appreciated.}, and  enable the resonances of \FII to go through the whole Trojan swarm. 
  The evolution of the planetary frequencies along the elliptical segment is represented in Fig.  \ref{fig:frec_plan} by red solid lines, while the green dashed curves show their evolution along the hyperbolic segment. Let us focus on the elliptic segment.
   Before going further, we have to mention that the quantity $\ineg{\alpha}{\beta}$ is an increasing function of $a_2$ as long as $\beta >0$. Even if it is more convenient to plot $\vert\ineg{\alpha}{\beta}\vert$ rather than its signed value, we have to keep in mind that this frequency is negative before an orbital resonance (i.e. when $a_2$ is smaller than the value required to be in MMR with Jupiter) and positive after. 
When $a_2$ evolves towards a MMR, the corresponding combination $\vert\ineg{\alpha}{\beta}\vert$ decreases towards zero. First, this frequency reaches values close to $8000\asay$ giving rise to the resonances $\nu \approx \ineg{\alpha}{\beta}$ of the second family.
Then, as $a_2$ increases, $\vert\ineg{\alpha}{\beta}\vert$ starts decreasing until it becomes comparable to a few $g$, where some resonance of family IV is encountered. When the planetary system crosses the resonance, the frequency  $\ineg{\alpha}{\beta}$, which is equal or at least very close to zero, suffers from chaotic variations due to the dynamical structures encountered inside the MMR (see Fig.  \ref {fig:res_plan}).  For the sake  of clarity,  the behavior of the frequencies inside  orbital resonances  is not reproduced in Fig.  \ref{fig:frec_plan}. This leads to gaps in the curves representing the frequencies. On both sides of these gaps, the singularities generated by the stable and unstable manifolds  of the resonance impose that the frequencies go to infinity.   

 The bottom frame of Fig.  \ref{fig:frec_plan} displays the values of the secular frequencies (red curves)  with respect to $a_2$. The negative frequency is $s_2$, while the two positive frequencies are $g_1$ and $g_2$ knowing that $g_1 <  g_2$. The general trend of the absolute value of the secular frequencies is to decrease when $a_2$ increases. This is merely due to the fact that the perturbations between the two planets decrease with respect to their mutual distance. But, as for the combinations of mean motions, singularities appear when the "separatrices" of the MMRs are reached. The increase in the secular frequencies is particularly striking on both sides of the 2:1 MMR. Here, as it can also be seen in a small neighborhood of the 7:3 and the 5:2 MMRs, $g_2$ is much more affected than the two other secular frequencies. More precisely, these three frequencies have to go to infinity, but the growth of the slope begins farther from the "separatrix" for $g_2$ than for $g_1$ and $s_2$.  As a result, we would expect to detect the influence of the secular resonances of \FIII like the  $g=(k+1)g_2-kg_1$ for at least  small values of $\vert k\vert$. But, as large values of $g_1$ and $g_2$ are necessary to reach these resonances, this phenomenon occurs only very close to the 2:1 MMR. As it will be shown in section \ref{sec:FII}, instability is so strong in this region that secular resonances cannot be isolated and clearly identified.

The behavior of the frequencies along the "hyperbolic segment" is qualitatively the same as along the "elliptic segment", except near the three dominant MMRs, where the singularities are shifted as predicted by Fig. \ref{fig:res_plan}. Indeed, the approach of a MMR is characterized by a sharp and sudden variation of the planetary fundamental frequencies corresponding to the singularity associated to the separatrix.  But the location of these asymptotes depends on the initial angles of the planets $(M,\omega,\Omega)$.
The consequences of this dependence on the initial phases will be studied in section \ref{sec:phases}.

\subsection{ Sweeping of  \FIV's resonances across the Trojan region.}
\label{sec:FIV}

\subsubsection{ Prediction of the location of \FIV's resonances }
\label{sec:predictionIV}
\begin{figure*} 
\includegraphics[width=22cm,height=8cm]{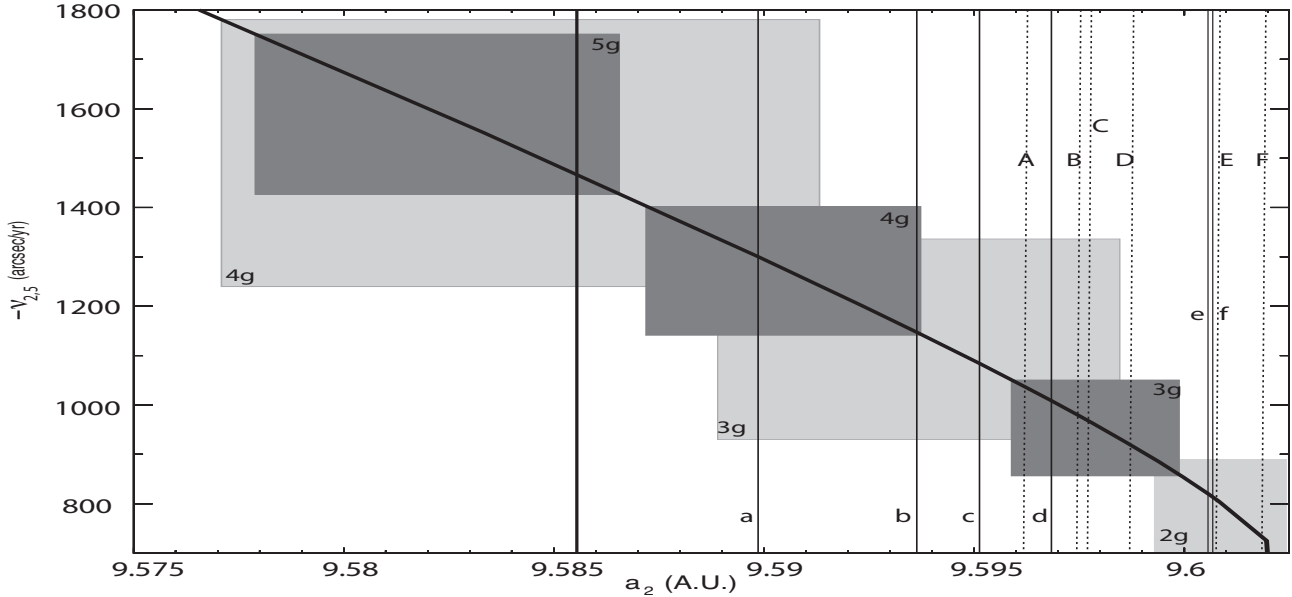}
\caption[]{ Crossing of the Trojan swarms by the resonances of \FIV.     The slanted bold black curve represents the values of $-\ineg25$ (\asay) versus $a_2$ (\au).  The grey rectangles indicate the values of $(a_2,\nu)$ for which a subfamily of  \FIV is inside the swarms. Light grey is used for $I^* = 2^{\circ}$ and dark grey for $20^{\circ}$. The vertical bold line near $9.585 \au$ corresponds to the present location of Saturn. The other lines indicate the values of $a_2$ used in the numerical simulations "a" to "f" ($I^* = 2^{\circ}$) and "A" to "F" ($I^* = 20^{\circ}$). See the text for details.  
}  
\label{fig:prediction_FIV}
\end{figure*}

 In this section we focus our attention on the close neighborhood of the 5:2 orbital resonance. Indeed, we have shown in \pu that the system is close enough to this resonance to give rise to narrow unstable regions resulting from \FIV . Even if these chaotic regions seem very thin, several resonances of this family  are involved in the low erosion process of Jupiter's Trojan swarms that was first mentioned by \citet{LevisonSS97}. 
Because the Trojans will encounter several resonances of this family between the current location of the system and the 5:2 MMR,  the study of this region enables us to describe and illustrate accurately the crossing of the Trojan swarms by the resonances of \FIV.  

The resonances of  \FIV, defined by formula (\ref{eq:FIV}), involve  the frequencies $\ineg25$, $g$, and the secular frequencies  of the planetary system. 
According to Fig.\ref{fig:frec_plan} and formula (\ref{eq:bounds}), the variations of $g_1,g_2,s_2$ are small with respect to those of $\ineg25$. Furthermore, the secular frequencies of the planets are small compared to $g$, except very close to the 2:1 MMR (see later). As a result, the location of \FIV depends mainly on the values of $\ineg25(a_2)$.
The variation of this frequency with respect to $a_2$ is plotted in figure \ref{fig:prediction_FIV}. More precisely, the bold curve of this picture represents the graph of the function $f$: $x \longmapsto -\ineg25(x).$ The X-axis corresponds to the initial values of $a_2$ (in \au) while the Y-axis is associated to the frequencies (in $\asay$).
 This curve  decreases very regularly, until $a_2$ reaches $9.602 \au$, where a sharp change in the slope indicates the beginning of the 5:2 MMR (see also Fig.\ref{fig:res_plan}, bottom-frame).   It is now straightforward to predict  the location of resonances belonging to \FIV in the Trojan phase space.  Indeed, as long as the right side of (\ref{eq:FIV}) is negligible with respect to $kg$,  the resonant condition is well approximated by the relation: 
 \be
 -\ineg25(a_2) \in \frac kj  \pi_g(\Theta_{I^*}).  
 \label{eq:approxFIV}
 \ee
  For given values of $j$ and $k$, the previous formula defines a frequency interval where the resonance is reached. Approximating the bold curve in Fig. \ref{fig:prediction_FIV} as smooth and monotonic, $f^{-1}$ maps this interval in another one of the X-axis\footnote{Strictly speaking, the function $f$ is not smooth on an interval, but, according to KAM theory,  at most on a Cantor subset of this interval, which, however, is nearly of unit measure. }, namely: $f^{-1}( (k/j) \pi_g(\Theta_{I^*}))$.
  The cartesian product of these two intervals, the first one on the frequency axis (Y) and the second one on the $a_2$ axis (X), defines a "resonant rectangle" displayed in gray in Fig.\ref{fig:prediction_FIV}, and denoted by $R^{k,j}_{I^*}$.  
  
   For the sake of clarity, our study is limited to the resonances defined by  $j=1$ and $k>0$ ($k<0$ on the other side of the MMR). 
Consequently, the rectangles $R^{k,1}_{I^*}$ will be denoted $R^k_{I^*}$.  According to (\ref{eq:bounds}) and (\ref{eq:approxFIV}), the location of the resonances of \FIV  depends on the initial inclination $I^*$. For this reason, we study sections of the  phase  space at two initial inclinations: $I^*=2\de$ and   $I^*=20\de$. The corresponding resonant rectangles are colored in light grey and dark  grey respectively.
The labels 5g, 4g, 3g and 2g in the bottom-left corner of the rectangles at $2\de$ and in their top-right corner at $20\de$ correspond to $k=5, 4, 3$ and $2$ respectively. 
  Fig.\ref{fig:prediction_FIV} shows that, for a given $k$, the size of the $R^k_{I^*}$ decreases with $I^*$, and that for a fixed  $I^*$, these rectangles are shifted rightwards as $k$ decreases. These two phenomena are due to the facts that the function $f$ is decreasing and concave, and that the width and the bounds of $\pi_g(\Theta_{I^*})$ decrease while $I^*$ increases.  One of the main consequences of these properties lies in the fact that \FIV enters the Trojan swarms for a value of $a_2$ which depends on $I^*$. Therefore, during a migration (slow monotonic variation of $a_2$ with respect to the time in the studied region), \FIV's  resonances  will sweep through the Trojans phase space in a way that depends on the inclination: $R^k_{2}$ being larger than $R^k_{20}$, it follows that the instability generated by \FIV will destabilize the swarms much more efficiently at low inclination than at higher inclination.  Moreover, instabilities associated to the $k$-subfamilies of \FIV corresponding to $R^k_{I^*}$, for different values of $k$, successively affect the swarms, a fact which further enhances the dependence of the degree of destabilization of the swarms on the value of the inclination  ${I^*}$. Finally, as Fig.\ref{fig:prediction_FIV} shows, the resonant rectangles can also overlap, ensuring the coexistence of at least two $k$-subfamilies in the same swarm. Such intersections and their induced dynamics are studied in the next section.

\subsubsection{ Crossing of the Trojan swarms by the resonances of \FIV.  }
\label{sec:crossingFIV52}

 In order to verify these predictions and to illustrate the geometry of the resonances of \FIV, we have integrated the $L_4$ Trojan swarm for several initial values of Saturn's semi-major axis $a_2$.
 The associated configurations of the planetary system  are represented in Fig. \ref{fig:prediction_FIV} by vertical lines, and the initial values of $a_2$ can be found in Tab. \ref{tab:FIV}.   The initial inclination of the Trojans is equal to $2\de$  for the solid lines, labeled  with small roman letters, and to $20\de$ for the broken lines labeled with capital letters. The present configuration of the Solar system, widely studied in \pu, is not discussed here, but is represented in Fig.\ref{fig:prediction_FIV} by the vertical bold  line located at $a_2\approx 9.5855 \au$. As $\ineg25$ is close to $-1467\asay$, this implies that $-\ineg25 \in 4\pi_g(\Theta_2)$ (light rectangle) and that $-\ineg25 \in 5\pi_g(\Theta_{20})$ (dark rectangle),  which is in perfect agreement with the results of \pu.

 The integer $j$ is already fixed to $1$, and from now on, we do not take into account the secular frequencies $s$ and $s_2$.
This omission is fully justified for small inclinations, and we will see later that even for an inclination equal to  $20\de$, the main resonances of \FIV are independent of $s$ and $s_2$. 
After this additional simplification, it is more convenient to rewrite relation (\ref{eq:FIV}) as:  
  \be
   g =  - \Frac{\nu_{2,5}}{k} + \Frac{k-3}{k} g_1 + \Frac{k'}{k}(g_2-g_1)
 \label{eq:famIV}
 \ee
 which satisfies relation (\ref{eq:FIV}) when $k =  3 - k_{g_1} - k_{g_2}$ and $k' = -k_{g_2}$.
This new simplified formulation of the resonance condition defining  \FIV enables us to exhibit some useful properties of these resonances.

 \begin{enumerate}
  \item For a given value of $a_2$, the right-hand side of (\ref{eq:famIV}) is a constant denoted $g_r$ (for simplicity, the parameters $k$ and $k'$ are omitted).
  \item Still assuming fixed $a_2$: as long as $k$ and $k'$ are not too big, the relation $\vert\ineg25\vert \gg \vert (k-3) g_1 + k'(g_2-g_1) \vert$ is satisfied (see Section \ref{sec:plan-freq}). Therefore, the resonance is reached for a value of $g$ very close to $-\nu_{2,5}/k$. Consequently, we can see $\FIV$ as being split in different subfamilies parametrized by $k$. In every subfamily, a single resonance is defined by the additional parameter $k'$.
  \item For a given $k$, the resonances of this subfamily are represented in the frequency space $\Theta_{I^*}$ by parallel planes which are separated by a distance of $(g_2-g_1)/k$. Moreover, these planes are arranged in increasing order, in the sense that $g_r$ increases with $k'$.  \end{enumerate}

According to property (ii), the resonant rectangles $R^k_{I^*}$ provide an approximation of the location of the $k$-subfamily which is valid as long as the combinations of the planetary secular frequencies involved in (\ref{eq:famIV}) are small with respect to $\ineg25/k$. 
Without going in too much detail, we can consider that these boxes are minimal in the sense that: at least one of the resonances of the $k$-subfamily belongs to $R^k_{I^*}$, and this resonance is always one whose dynamical influence is the largest. More precisely, except for $k=3$, $g=-\nu_{2,5}/k$ does not satisfy equation (\ref{eq:famIV}), and consequently does not correspond to any resonance of \FIV. But its closest resonant value, namely $-\nu_{2,5}/k + (k-3)g_1 /k $, is reached for $k'=0$ in formula (\ref{eq:famIV}). If this resonant value does not belong to the box $R^k_{I^*}$, one of the two frequencies $- \nu_{2,5}/k + (k-3) g_1/k   \pm (g_2-g_1)/k$ does. Furthermore, the resonance associated to $k'=0$, or one of its two closest neighbors, plays a central dynamical role in the subfamily because its order is minimal (here we use the classical definition of order, that is:   $\vert k\vert +\vert k - k' -3 \vert + \vert k'\vert$), implying that its width are larger than the widths of all other resonances in the same family. 
 Thus the condition $g=-\ineg25 /k$ suffices to predict approximately the value of $a_2$ at which a major resonance of \FIV will influence the Trojans.

\begin{figure} 
\includegraphics[width=8cm]{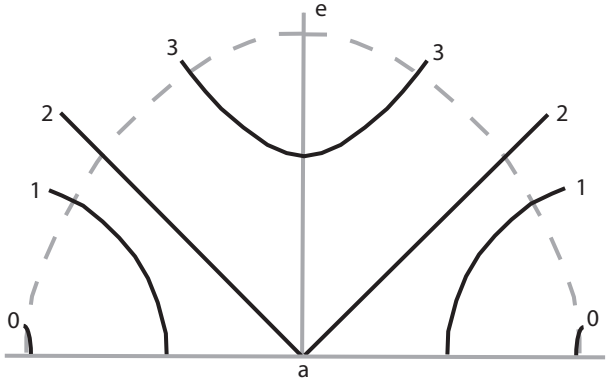}
\caption[]{ Schematic view of the displacement of a resonance of \FIV through the Trojan swarm. The X-axis represents the initial value of Trojan's semi-major axis. The initial eccentricity is associated to the Y-axis. The dashed curve approximates the limit of the stability region. The black curves correspond to the location of the same resonance for different values of Saturn's initial semi-major axis $a_2$. The larger the label, the larger $a_2$.}
\label{fig:schema}
\end{figure}

Before going further, we will describe schematically the motion of a single resonance of \FIV during its crossing of the co-orbital region. 
Fig. \ref{fig:schema} represents a section of the corresponding  phase space in the $(a,e)$ plane where the other initial elliptic elements are fixed.
The horizontal straight line is the $a$-axis where $e= e_1^{(0)}$ while the vertical grey line is the $e$-axis where $a = a_1^{(0)}$. The two symmetrical sides of the phase space section are represented here ($a>a_1^{(0)}$ and $a<a_1^{(0)}$).  Indeed, as it is explained in \pu,  the  phase space of Jupiter's Trojans is symmetric. More precisely, we can clearly observe two different types of symmetries. 
The first one, which was already visible in \cite{MiBeRo2001,NeDo02}, is a symmetry with respect to a curve that is close to the straight line $a=a_1^{(0)} \approx 5.2035\, \au $ 
(i.e., the initial semi-major axis of Jupiter), and tangent to it at $L_4$. 
The second symmetry is with respect to a curve  close to the axis $e=e_1^{(0)} \approx 0.0489$. Moreover, considering quasiperiodic trajectories,  the fundamental frequencies corresponding to a given initial condition  and the ones corresponding to one of its two symmetric points are the same. These frequencies parametrize the KAM torus on which the given trajectories lie. 
This does not mean that the two corresponding trajectories are the same, but that they lie on the same invariant torus. From the dynamical point of view, these  trajectories are equivalent. 
These symmetries point out the fact that there are manifolds (even close to $L_4$) on which the  frequency map is degenerated  (see \cite{GaJoLo2005}), and they allow us to restrict the sample of initial conditions to the subset $\{(a,e);\, a \ge a_1^{(0)}  ,e \ge e_1^{(0)}  \}$. 

\begin{figure*}
\begin{tabular}{cc}
\includegraphics[width=8cm,height=10cm]{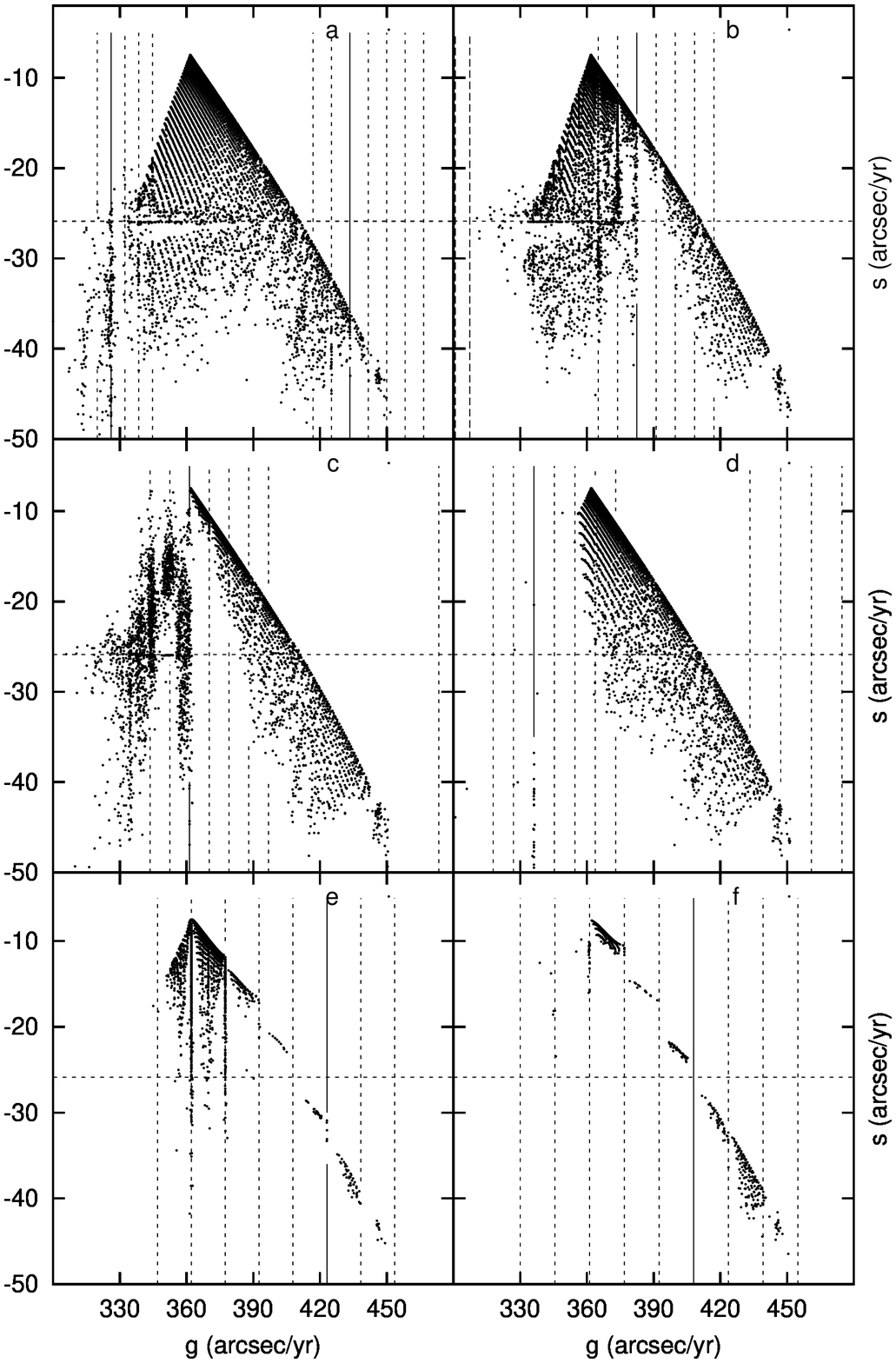} &
\includegraphics[width=8cm,height=10cm]{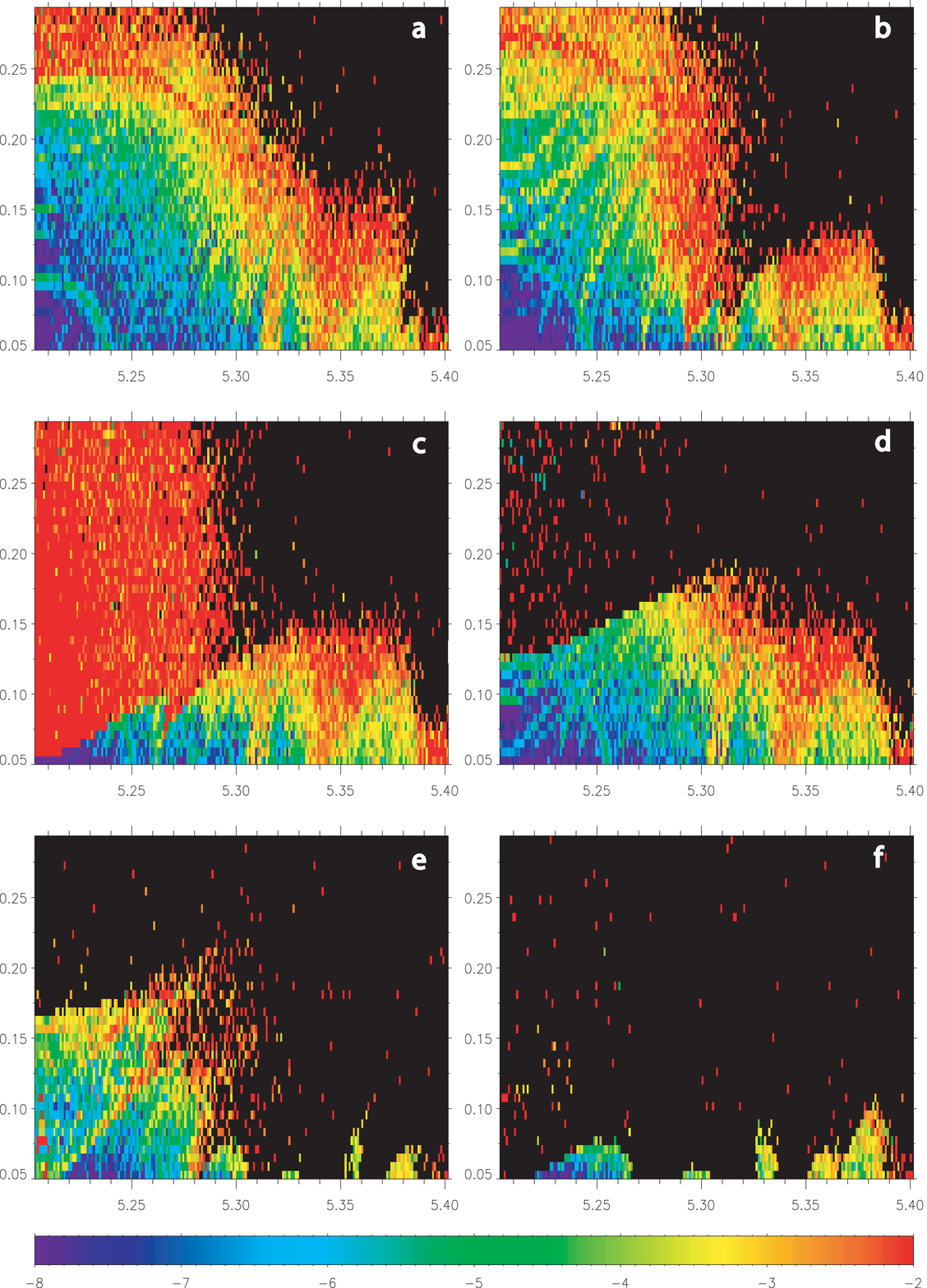}\\
\end{tabular}
\caption[]{Destabilization of the Trojan swarms by  the resonances of  \FIV at $I^* = 2\de$. Left block: Dynamical maps in the frequency space (projection on the $(g,s)$ plane).  The vertical lines show the predicted location of the resonances of \FIV involved in the dynamical process (see Table \ref{tab:FIV}), while the horizontal line emphasizes the secular resonance $s = s_6$.  Right block: Dynamical maps in the action space  (the initial conditions of the fictitious Trojans are chosen in the $(a,e)$ plane, the four other elliptic elements being fixed). The diffusion rate defined by  $\log_{10}\vert\frac{\nu^{(1)}-\nu^{(2)}}{\nu^{(1)}}\vert$ is coded by colors that vary  from $-8$ (stable) to $-2$ (strongly chaotic).  Each block is split in six panels labeled from "a" to "f" corresponding to the numerical simulations described in section \ref{sec:crossingFIV52}. See the text for more details.
}
\label{fig:frec_FIV_I02}
\end{figure*}

\begin{figure*}
\begin{tabular}{cc}
\includegraphics[width=8cm,height=10cm]{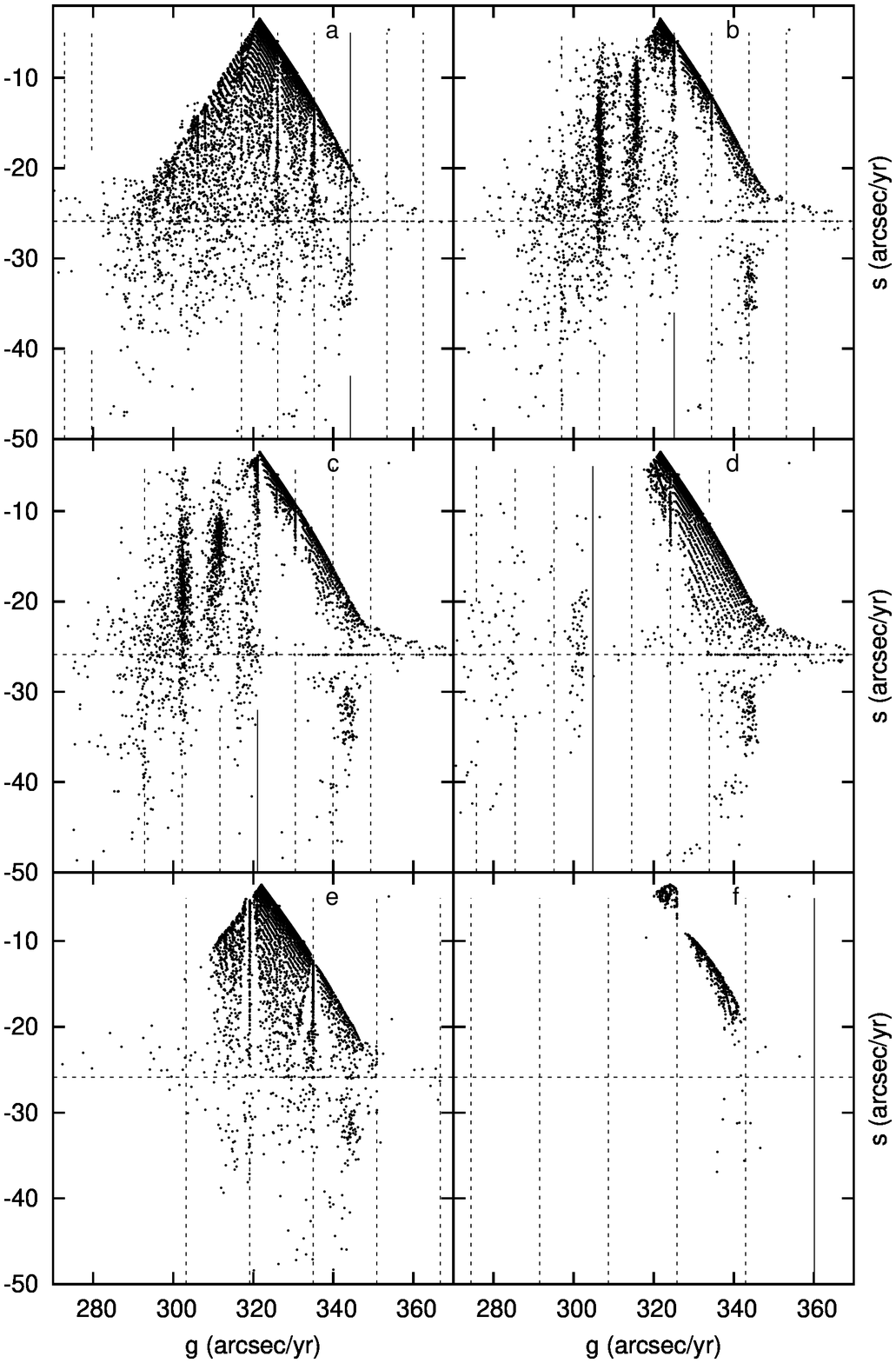} &
\includegraphics[width=8cm,height=10cm]{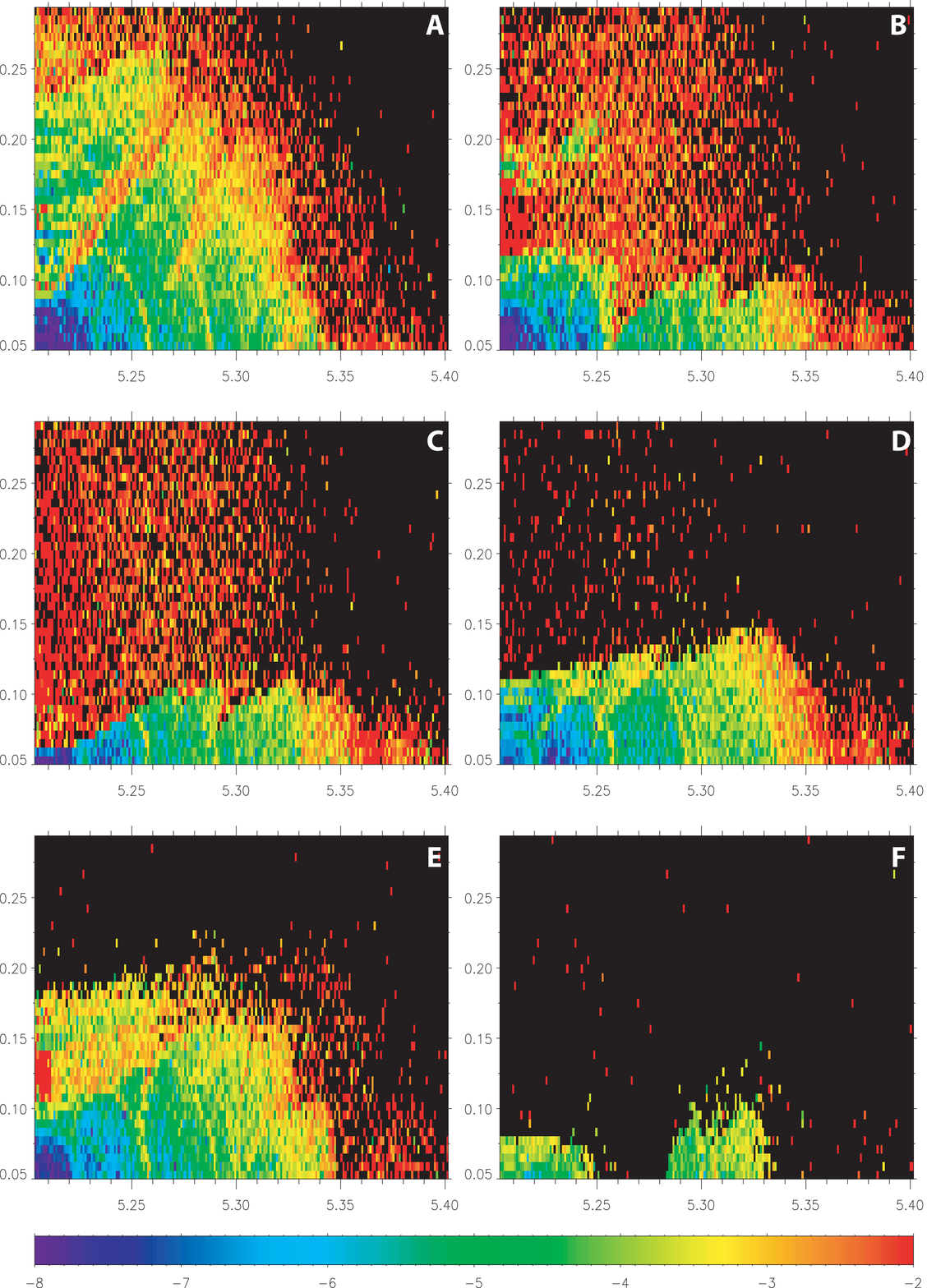}\\
\end{tabular}
\caption[]{Same as Fig.\ref{fig:frec_FIV_I02} for $I^* = 20\de.$}
 \label{fig:frec_FIV_I20}
\end{figure*}

The dashed grey curve gives the lower bounds of the strongly unstable region (the real form of this boundary corresponds to the limits of the dark domains in all the right panels of Figs \ref{fig:frec_FIV_I02} and  \ref{fig:frec_FIV_I20} which shall be discussed in detail below).  Let us assume that we start the evolution of the planetary system with an initial value of $a_2$ locating Saturn between Jupiter and the 5:2 MMR, in such a way that  $a_2$ increases during the migration.
As the planetary system gets closer to the 5:2 MMR, the frequency $-\ineg25/k$ decreases towards zero ( $\ineg25$ is negative in this region).  Consequently the resonant frequency $g_r$, associated to the $k$-subfamily decreases too, and the first contact between the Trojans and this resonance arises at the two black segments labeled with $0$, where $a$ reaches its lowest and greatest value. Then, as $g_r$ keeps decreasing (and $a_2$ keeps growing), the resonance goes towards the center (black curve 1) to reach $L_4$ (label 2). Here, the two separated branches merge together to give a single curve. After the resonance travels through the vicinity of $L_4$,  it moves towards higher eccentricities (label 3) and leaves the Trojan swarm through the secular resonance $s=s_2$ (at least for low to moderate initial inclinations).

This schematic view  is essential to understand the evolution of the dynamics of Trojan swarms during planetary migration. Such  evolution is presented in Figs \ref{fig:frec_FIV_I02} and \ref{fig:frec_FIV_I20}. These figures show the sweeping of the co-orbital region, at $I^* = 2\de$ and $I^*= 20\de$ respectively, by the resonances of \FIV. 
Figs \ref{fig:frec_FIV_I02} and \ref{fig:frec_FIV_I20} are composed of two blocks. The right block corresponds to dynamical maps of the domain $D_{2\de}$ (Fig \ref{fig:frec_FIV_I02}) and $D_{20\de}$  (Fig \ref{fig:frec_FIV_I20}). As mentioned in section \ref{sec:model}, we use the relative change of the frequency $\nu$ given by  $\sigma_{\nu} = (\nu^1-\nu^2)/\nu^1$, called diffusion index, as an indicator of the regularity of the motion. In this formula, $\nu^1$ is the libration frequency computed on the first 5 My, while $\nu^2$ is calculated on the following 5 My. 
Figs.~\ref{fig:frec_FIV_I02} and \ref{fig:frec_FIV_I20} (right block) show dynamical maps in the action-like space (here $a$ and $e$). A color is assign to each fictitious Trojan, coding its diffusion index. 
The color scale ranges from blue, which corresponds to stable regions ($\sigma_{\nu}<10^{-6}$), to red for very chaotic regions ($\sigma_{\nu}>10^{-2}$). In black, we display the particles that have been ejected from the Trojan swarms during the integration (10 My).
The left block is the corresponding view in the frequency space. The frequency map ${\cal F}$ establishes the correspondence between these two blocks which are dynamically equivalent. 
Consequently, figures~\ref{fig:frec_FIV_I02} (left) are made of an union of curves (more or less smooth)  which are the images of the lines $e_0={\rm constant}$ by ${\cal F}$. 
In the stable regions, the frequency map is very smooth implying the smoothness of the above mentioned curves (this is typically the case for $s> -20\asay$ in Fig.  \ref{fig:frec_FIV_I02}.a and for  $s> -15\asay$ in Fig.  \ref{fig:frec_FIV_I02}.b). In contrast, in chaotic regions, ${\cal F}$ is singular, and the considered curves lose their smoothness. They are in fact disconnected and the points composing these curves seem to scatter as in Fig.  \ref{fig:frec_FIV_I02}.a for $s< -30\asay$ (see \cite{La99}  for more details).
Each block is made of six panels, labeled from 'a' to 'f' at $I=2\de$ and 'A' to 'F' at $I=20\de$, associated to different values of $a_2^{(0)}$. These values are represented in Fig. \ref{fig:prediction_FIV} by vertical lines crossing the grey boxes.   These values are chosen such that from 'a' to 'f' (resp. 'A' to 'F') Saturn's semi-major axis increases, allowing the $k$-subfamilies of \FIV to cross the Trojan swarms following the rules established in section \ref{sec:predictionIV}. 
The resonances of \FIV are clearly visible in the frequency maps (left blocks) as vertical structures made of gaps or of accumulations of dots.  These are enhanced by vertical dashed lines for $k' \neq 0$  and by solid lines for $k'=0$, where $k'$ is the integer which appears in formula (\ref{eq:famIV}) as the parameter of the $k$-subfamily.  On the action side (right blocks), it is not so easy to identify these resonances. Although having the typical shape drawn in Fig.\ref{fig:schema}, confusions are always possible since the elements of \FII also have a quite similar form.
The main resonances of \FIV, which are drawn in Figs \ref{fig:frec_FIV_I02} and \ref{fig:frec_FIV_I20}, are gathered together in  Tab. \ref{tab:FIV}. The initial values of Saturn's semi-major axis are listed in its  second column. The corresponding labels are in the first column. The two last columns correspond to the values of the integers $k$ and $k'$ appearing in formula (\ref{eq:famIV}) and which define the resonances of \FIV associated to the 2:5 MMR.

\begin{table}
\caption[]{ Resonances of \FIV represented in Figs \ref{fig:frec_FIV_I02} and \ref{fig:frec_FIV_I20}. The two first columns indicate the label of the corresponding panel and the initial value of $a_2$ used for the associated numerical simulation. The two last columns give the values of the integers $k$ and $k'$ defining jn formula (\ref{eq:famIV})  the resonances represented by vertical lines on each panel of Figs \ref{fig:frec_FIV_I02} and \ref{fig:frec_FIV_I20}. }
\begin{tabular}{clcc }
\hline
 Label   & $a_2$ (\au) & $k$ & $k'$ \\
\noalign{\smallskip}
a & 9.58986 & 4 & $ -1, \cdots,  \phantom{0}3$ \\	
-  & -               & 3 & $ -2, \cdots,  \phantom{0}4$ \\	
b & 9.59364 & 4 &  $ 2, 3$ \\	
-  & -               & 3 &  $ -2, \cdots,  \phantom{0}4$ \\	
c & 9.59513 &  3 &  $ -2, \cdots,  \phantom{0}4$ \\
-  & 9.59513 &  2 &  $ -5$ \\
A & 9.59630  & 4 & $ 2, 3$ \\
- & 9.59630  & 3 & $-3, \cdots,  \phantom{-}2$ \\
d & 9.59684 &  3 &  $ -2, \cdots,  \phantom{-}4$ \\		
-  & -               &  2 &  $ -5, \cdots,  -2$ \\	
B & 9.59756  &  3 & $ -3, \cdots,  \phantom{0}3$ \\
C & 9.597816  &  3 & $ -3, \cdots,  \phantom{0}3$  \\
D & 9.59882  &  3 & $ -3, \cdots,  \phantom{0}3$ \\
e & 9.60004 &  2 & $ -5, \cdots,  \phantom{0}2$ \\		
f  & 9.60057 &  2 & $ -5, \cdots,  \phantom{0}3$ \\
E & 9.60088 & 2 & $-6, \cdots, -2$ \\
F & 9.60197  & 2 & $-5, \cdots,  \phantom{0}0$ \\		
\end{tabular}
\label{tab:FIV}
\end{table}

The secular resonance $s=s_2$ is also easily identifiable in the left block by a horizontal line corresponding to an accumulation of dots at $s = -26\asay$ surrounded by gaps. This resonance splits the frequency space in two different regions: the part located above the resonance, which is globally regular assuming that the planetary system is not too close to the 5:2 MMR (it is true at least between the Saturn's present location $a_2 = 9.5855 \au$ and  $9.595 \au $ simulation $c$), and the part lying below this resonance which is strongly chaotic.  This secular resonance appears in the actions space (right block) as the red arch crossing the vertical axis at about $e= 0.25$ and the horizontal one at $a = 5.35 \au$.  As mentioned above, Fig. \ref{fig:frec_FIV_I02}.a  reveals that the domain located inside this arch (blue to green color) is much more regular than the outside part (orange to black). For small eccentricities, $s=s_2$ overlaps with the \FI resonance defined by formula (\ref{eq:FI}) with $i=1$ et $j=13$ (red V-shape around $a = 5.35 \au$) , generating this large chaotic region around $5.34 \au$. \FI is also acting around $5.4 \au$ where the narrow red strip corresponds to  $i=1$ and $j=14$. But, as in this experiment the semi-major axis of Saturn varies only very little  (about one hundredth of $\au$ (see Tab. \ref{tab:FIV})) only the resonances of \FIV move across the Trojan swarm, while those belonging to the three other families are practically fixed.  This makes it easier to recognize the resonances of \FIV. 

  According to \pu, the present system (vertical bold line  in  Fig. \ref{fig:prediction_FIV}) is such that the $4$-subfamily of \FIV is acting on low inclined Trojans. For slightly larger $a_2$, both the $4$-subfamily  and the $3$-subfamily are involved in the Trojans' dynamics.  In Fig.\ref{fig:frec_FIV_I02}.a, the $4$-subfamily is exiting the phase space (the corresponding resonances are located around $g=330\asay$), while the $3$-subfamily penetrates inside. For $k=4$, the subfamily mainly appears above the arch $s=s_2$ at high eccentricities ($e>0.25$), and strengthens the instability induced by the secular resonance.  The $3$-subfamily appears at the opposite side of the frequency space  and is very difficult to identify in the action space because it is located in a region already filled by the resonances of \FI mentioned above (red V-shape surrounding $a_2 = 5.35 \au$). It is worth mentioning that the narrow gap at $a =5.38 \au$ in Fig. \ref{fig:frec_FIV_I02}.a is generated by the \FIV resonance defined by $k=3$ and $k'=1$.  This resonance will play a major role in the next experiment. 
     When $a_2$ is slightly larger, as in simulation 'b' ( $a_2 = 9.59364 \au$), the situation is much more interesting. Indeed, the subfamily $k=3$ begins to reach the central region, generating strong instability associated to a rapid escape of numerous Trojans, particularly for $a>5.31\au$. 
 Another important point lies in the fact that this subfamily encounters elements of \FII. This overlap generates the deep gap located at $5.31\au$.    The involved resonances of \FII are defined by the multiplet $(\alpha,\beta, i, j, k, l, l_s) = (1,2,5,2,0,0,0)$ in formula (\ref{eq:FIIp}) and are identifiable  in Fig. \ref{fig:frec_FIV_I02}.a by the nearly vertical orange structure between $5.31$ and $5.32 \au$ (Its location is practically unchanged from  Fig. \ref{fig:frec_FIV_I02}.a to Fig. \ref{fig:frec_FIV_I02}.f).
  
 In panel 'c', the $3$-subfamily continues to cross the Trojan swarm, and now occupies its core. The resonance associated to $k'=0$ is now very close to $L_4$ (curve 2 in Fig \ref{fig:schema}) while previous ones ($k'<0$) are on the other side of $L_4$ (curve 3 in Fig \ref{fig:schema}).
 The overlap of these resonances (for $k'\neq 0$) generates the huge red region above the resonance $k'=0$, while the other part of the phase space begins to be a bit more regular. 
 In panel 'd', the $3$-subfamily moves outwards but its resonances are still acting for $e>0.13$. The sharp transition between the black region and the blue one is due to the resonance defined by $k=3$ and $k'=2$. Below this resonance, the size of the stable region (blue to green) is increasing again. Let us remark that the effect of the $2$-subfamily (right side of the frequency map) is not yet noticeable. 
 Finally, when the $2$-subfamily penetrates deep inside the Trojan swarm, the dynamics becomes much more chaotic. 
 In panel 'e', a huge number of Trojans escapes the co-orbital region in less than ten million years, while in 'f', only small islands of temporary stability remain. The comparison between 'f'-left and 'f'-right shows that each island is separated by two consecutive resonances of the $2$-subfamily. This is possible because the distance between two resonances of the previous kind is larger than the one separating the resonances  of the $3$ and $4$-subfamilies.  The simulation 'f' is the last one of the sequence. Indeed, if Saturn get closer to the 5:2 MMR, the entire Trojan swarm is rapidly cleared. At $9.602 \au$, (see Fig. \ref{fig:prediction_FIV}), the planetary system is getting closer to the MMR, and the frequency $-\ineg25$ decreases abruptly. During this decrease, the $1$-subfamily is reached, but at the same time, the planetary system crosses the 'separatrix' of the 5:2 MMR. The chaos induced by the separatrix crossing combined with the resonances of \FIV make the whole Trojan swarm strongly unstable. We will see in section \ref{sec:FII} that this phenomenon stops when the planetary system is inside the stable domain of the libration area associated to  the 5:2 MMR.
 
 At  an initial inclination of $20\de$, the mechanism is practically the same, but a bit delayed. Indeed, in simulation 'c' ($I^*=2\de$) and 'C'   ($I^*=20\de$) the dynamical situation is basically the same, but 'c' occurs at $a_2=9.59513 \au$ and 'C' at $9.597816 \au$. This is a direct consequence of the fact that the upper bound of $\pi_g(\Theta_{I^*})$ is a decreasing function of $I^*$.    This may have implications on the final distribution of the Trojans versus  inclination, as we will see in the end of the present section. 
Looking at Fig. \ref{fig:frec_FIV_I20} (left block), one realizes that the main chaotic structures are vertical. This confirms the choice, which could seem arbitrary, that we have made in formula (\ref{eq:famIV}).  Indeed, the resonances of \FIV which generate the most chaotic behavior do not depend on  $s,s_j$.  Conversely, these secular nodal frequencies are involved in secular resonances.   In addition to the resonance $s=s_2$, which is still visible at $20\de$, two other resonances can be seen in Fig. \ref{fig:frec_FIV_I20} (right block). These structures form the two yellow thin arches (particularly in 'A') cutting the X-axis at $5.26\au$ for the  $ 3s-s_2-2g_1 = 0$ (inner arch), and $5.29\au$ for the $2s-3g_1+g_6 = 0$ (outer arch).
As in Fig. \ref{fig:frec_FIV_I02}, simulations 'A' to 'F' show the shifting of the resonances of \FIV. In 'A', the $4$-subfamily is exiting the phase space while the $3$-subfamily has already reached its center. The two resonances in both sides of $L_4$ are defined by $k=3$ and $k'=-3,-2$. They are associated to the two quasi-straight lines displayed in yellow-orange colors, which cross the two above-mentioned resonant arches.
In 'B', the resonances are shifted leftwards (frequency space) and their overlap generates global chaos above $e=0.12$ (location of the resonance $k=3$, $k'=-1$). Then, the resonance $k'=0$ reaches $L_4$ (in 'C') and generates the sharp transition between the blue region and the red region, while the size of the regular region keeps shrinking.
Next, as the $3$-family moves outwards, the stable area begins to increase up  to simulation 'E' where resonances of  the $2$-subfamily with large $\vert k\vert$ penetrate the phase space to finally  eject most of the Trojans in 'F'.  
As in simulations performed with an initial inclination of $2\de$, only small islands of temporary stability remain and are surrounded by gaps generated by the resonances of the $2$-subfamily.
The increase in the size of the stable region between the crossing of two consecutive subfamilies of \FIV is something that we have not observed at low inclinations.  The explanation of this difference is very simple. We know from section \ref{sec:predictionIV}, that the rectangles  $R^k_{2\de}$ are larger tan the $R^k_{20\de}$ and that the boxes $R^k_{2\de}$ overlap while the $R^k_{20\de}$ do not. It turns out that in the gap lying between $R^k_{2\de}$ and $R^k_{20\de}$, the Trojan swarm recovers, at least partially, its stability.
This phenomenon can be also observed in Fig. \ref{fig:prediction_FII}.b around $a=9.6 \au$, where the fraction of ejected Trojans decreases suddenly from $0.8$ to less than $0.6$ to finally jump to $0.9$ and more (see section \ref{sec:FII} for more details).

To complete this section, let us point out that the above-mentioned evolution may lead to a  lack of weakly inclined Trojans. Indeed, if we assume that the migration stopped close enough to the MMR, the sweeping of the Trojans would have taken  place for small inclinations while it wouldn't have begun for higher inclinations. Unfortunately, in the Solar system, Jupiter and Saturn are too far from the 5:2 MMR to have generated this phenomenon; except if this system was,  in the past, closer to the resonance than it is now. But it does not seem to be a realistic scenario. 

\subsubsection{ \FIV associated to other orbital resonances  }
 \label{sec:crossingFIV73}
 \begin{figure}
\includegraphics[width=6cm,height=6cm]{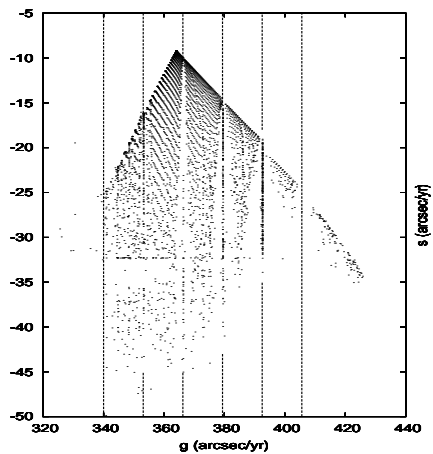}
\caption[]{The 3:7 MMR and its \FIV: $\Theta_{2}$ projected on the $(g,s)$ plan for $a_2= 9.1833 \au$.
 From right to left the six verticals lines correspond to the resonances $2g = -\nu_{3,7} - 2g_1 + k^{'}(g_2-g_1)$ for $-k \in \{1,2,3,4,5,6\}$. The frequency unit  is the $\asay$. } 
\label{fig:2g37}
\end{figure}

Since the mechanism studied in section \ref{sec:crossingFIV52} is very general, resonances of \FIV should take place in a small neighborhood of each MMR. But in some cases, the corresponding region is so chaotic that no resonance can be accurately identified. This is particularly true for the 2:1 orbital resonance, where its merging with \FII generates a huge unstable zone (see section \ref{sec:FII}).

On the other hand, the resonances of \FIV associated to the 7:3 MMR can be easily  detected, essentially because they generate moderate chaos.
We consider here that the initial semi-major axis of Saturn is equal to $a_2= 9.1833 \au$. Fig. \ref{fig:2g37} displays the projection of the Trojans frequency domain $\Theta_{2}$ on the plane of coordinates $(g,s)$. 
  For this planetary configuration, which is very close to the 7:3 MMR, the combination of proper mean motion  $\nu_{3,7} = 3n_1 - 7n_2$ is close to $-423 \asay$, implying the relation $-\nu_{3,7}/2 \in \pi_g(\Theta_2)$. Consequently, several resonances of the $2$-subfamily, defined by the relation:
   \be
 2g = -\nu_{3,7} - 2g_1 + k^{'}(g_2-g_1),
 \ee
 are present in the Trojans' phase space.  Fig. \ref{fig:2g37} shows those resonances for $k'\in\{-6,-5,-4,-3,-2,-1\}$.
 Let us notice that the horizontal line, corresponding to the secular resonance $s=s_2$, is shifted downward in comparison with Fig.  \ref{fig:frec_FIV_I02}. Indeed, $s_2$ was equal to about $-26\, \asay$ in the simulations of section \ref{sec:crossingFIV52} while here, its value is close to $-32\, \asay$.

\subsection{ Role of the secondary resonances of \FII}
\label{sec:FII}

\begin{figure*}
\includegraphics[width=12cm,angle=270]{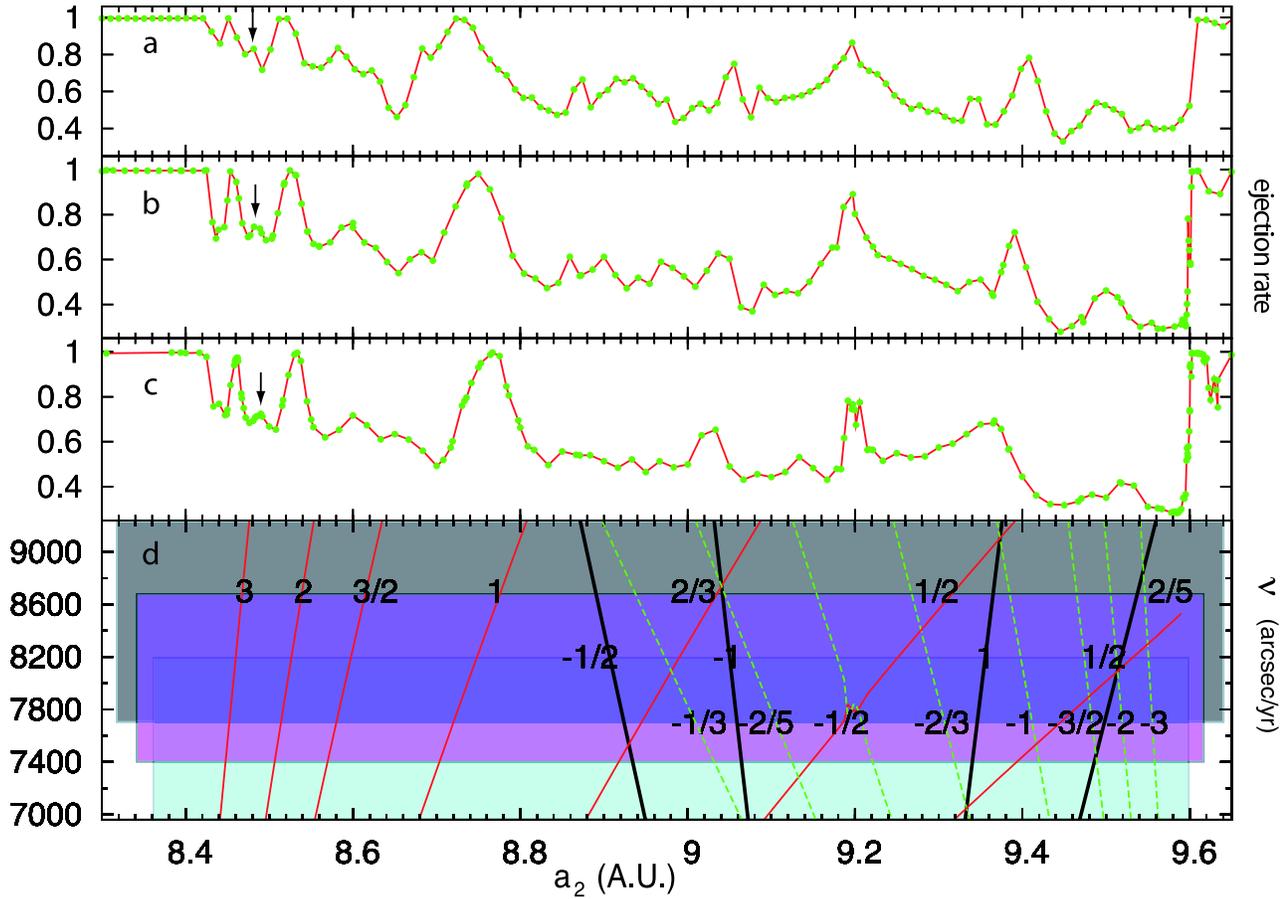}
\caption[]{ Chaos generated by the resonances of \FII: 
on the bottom frame (d),  the frequencies $\nu_{1,2}$,$\nu_{3,7}$,$\nu_{2,5}$ and some of their rational multiples are plotted versus $a_2$. The rational numbers written on the curves $(i/j)\nu_{(\alpha,\beta)}$ correspond to the value of $i/j$, the values of $(\alpha, \beta)$ being equal to $(1,2)$ for the red curves, $(3,7)$ for the black bold ones and $(2,5)$ for the green dashed curves. The horizontal grey, pink and blue strips indicate the reachable values of $\nu$ in $\Theta_2$, $\Theta_{20}$, and  $\Theta_{30}$. The three upper plots represent the relative number of ejected trojans $n_{eject}$ (escape rate) during our numerical integration (section \ref{sec:model}) of respectively $\cD_{2\de}$ (panel c), $\cD_{20\de}$ (b)  and $\cD_{30\de}$ (a) for $a_2\in[8.3,9.65]$ \au. The arrows indicate the location of the secular resonance $g = 3g_2 - 2g_1$. }
\label{fig:prediction_FII}
\end{figure*}

\begin{table}
\caption{Secular resonances: values of $a_2$ (in \au) for which the three first resonances of the form $g=g_2+k(g_2-g_1)$  enter the intervals  $\Theta_2$, $\Theta_{20}$, and $\Theta_{30}$ for the "elliptic section"  $(M_2=340.04\de)$.
}
\begin{tabular}{|c|c|c|c|}
\hline
$I^*$   &  $g=g_2$  & $g=2g_2-g_1$   &  $g=3g_2-2g_1$\\
& & & \\
$2\de$   &  $<8.39$  &  $[8.41,8.425]$  &  $[8.43,8.455]$ \\
$20\de$   & $<8.395$  &  $[8.425,8.43]$  & $[8.45,8.46]$\\
$30\de$   &  $<8.4$  &  $[8.435,8.44]$  & $[8.47,8.475]$\\
\end{tabular}
\label{tab:res_sec}
\end{table}

Up to now, we focussed on the resonances of \FIV to clearly illustrate the displacement of instabilities inside the Trojan swarms during planetary migration. But investigations regarding this family, which only acts on a relatively small neighborhood of its associated MMR, confine our study to specific periods when the planets are close to MMRs.  In this section we will adopt a more global point of view by trying to find out, in the most exhaustive possible way, which are the resonances able to affect Trojan swarms during a migration phase. As mentioned in section \ref{sec:model}, we limit the migration path to the domain $a_2  \in [8, 9.65] \au$, which locates Saturn between the 2:1 and the 5:2 MMRs. Our goal can be reached in, at least, two different ways. The first one consists in the prediction of these events knowing on one hand the behavior of the planetary frequencies, and on the other hand the four families of resonances generating chaotic behaviors. 
The second way relies on numerical simulations of the same kind as the ones presented in section \ref{sec:crossingFIV52} and \ref{sec:crossingFIV73}.
We have followed independently these two different paths and have found results in very good agreement. These results are presented in Fig \ref{fig:prediction_FII}. Among the numerous data produced by these numerical simulations, we only plot the ejection rate, that is:  the number of escaping Trojans divided by the amount of bodies present in the initial  population.
 Fig.  \ref{fig:prediction_FII} is split in four panels. The bottom frame (\ref{fig:prediction_FII}.d) shows the predicted location of the resonances generating instabilities, described in detail later on. The three other panels display the depletion rate (Y-axis) with respect to the initial semi-major axis of Saturn (X-axis, in \au). They correspond to simulations with initial inclination equal to $I^*=2\de$ in Fig.\ref{fig:prediction_FII}.c , $I^*=20\de$ (\ref{fig:prediction_FII}.b), and $I^*=30\de$ (\ref{fig:prediction_FII}.a).  On these depletion curves, the succession of steep peaks of ejection (local maximum of ejection) and deep troughs of stability (local minimum of ejection) provide useful indications on the global dynamical behavior of the Trojan swarms along the migration path.

Before going further, let us notice that, in the present Solar System, the Jovian Trojans are located in a trough of strong stability ($a_2 = 9.5855\au$). 
It may be amazing that the migration would stop where the stability of the Trojan swarms is maximal. But in fact, other planetary configurations leading to stable swarms exist. According to  Fig.  \ref{fig:prediction_FII}.a-c, the most stable regions lie between $9.43 \au$ and the left edge of the 5:2 MMR.  In particular,   $a_2=9.44 \au$ provides a nice solution to this problem which is perhaps the most stable that we have explored, particularly at an initial inclination of $20\de$ and $30\de$.

\begin{figure} 
   \centering
   \includegraphics[width=9cm]{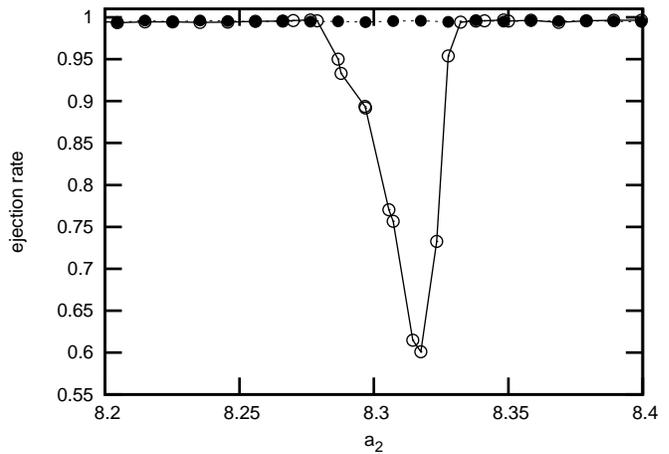} 
   \caption{Two different crossings of the 2:1 MMR. The dashed line with black circles corresponds to the "elliptic section" already presented in Fig. \ref{fig:prediction_FII}. Along this line, the diffusion rate (vertical axis) is always close to 1. On the contrary, the solid line which crosses the 2:1 MMR along the line defined by $M_2 = 316^{\circ}$ indicates the presence of most stable Trojan swarms for $a_2 \in [8.28:8.33]\, \au$ (see the text for more details).   }
   \label{fig:zoom21}
\end{figure}

The ejection peaks present in Fig.  \ref{fig:prediction_FII} are obviously associated to resonances.We will show in this section that, except three peaks, the other ones are generated by resonances of \FII. 
 Let us first consider these three particular cases. Those unstable regions are indirectly generated by the three main MMRs present in the studied domain: the 2:1 MMR for $a_2 < 8.4 \au$ (in fact this is more a plateau than a peak, in this whole region the depletion rate is about $100\%$), the 7:3 MMR at $a_2 = 9.2\au$ and the 5:2 MMR for $a_2$ around $9.6\au$ and more.  As mentioned in section  \ref{sec:plan-freq}, both sides of these peaks are generated by
 action of \FIV conjugated to the chaos introduced by the planets crossing the separatrices of the MMRs. When the planetary system is deep inside the  7:3 and more importantly  5:2 MMRs, the depletion rate decreases suddenly, which suggests the existence of regions harboring stable Trojans. This is especially striking at $I^* = 2\de$.   This phenomenon has been pointed out in \cite{MaScho2007} where authors find long-lived Trojans when Jupiter and Saturn are in 2:1 orbital resonance. We do not observe the same stability in this resonance, as we can see in Fig.\ref{fig:prediction_FII} where the ejection rate is always close to $100\%$. 
These different behaviors can be explained quite easily. Indeed, Trojan stability depends strongly on the location of the planetary system with respect to the different topological structures of the considered MMR. If the planets inside the resonance are close a separatrix (hyperbolic manifold) of one of these structures, the chaos induced by the planets may destabilize the whole Trojan swarm. On the contrary, if the planets lie in a stable region corresponding for example to an elliptic libration centre, their motion is in general close to quasiperiodic. Consequently, the Trojans do not suffer from chaos induced by the planetary system. But the motion of the Trojans is not necessarily regular. Secondary resonances between $\nu$ and the libration frequency associated to the 1:1 MMR can arise, leading to the ejection of numerous Trojans. Without studying the MMRs one by one, we cannot know which case corresponds to a given situation. We can only mention that, around $9.63\au$, where a region of relative stability takes place inside the 5:2 MMR, the angle  $2\lambda_1 - 5\lambda_2  +  3\varpi_2$ is librating, and according to Fig. \ref{fig:res_plan}, our 'elliptic migration path' crosses some kind of libration center (one of the most stable regions encountered in the section of the phase space).  On the contrary, when Jupiter and Saturn are in 2:1 MMR ($a_2 \leq 8.4\au$), almost all Trojans are ejected from the co-orbital zone. This is  because our migration path always stays quite far from the libration center of the 2:1 orbital resonance (see Fig. \ref{fig:res_plan}). Fig.\ref{fig:zoom21} shows a very different behavior when the segment of initial conditions crosses the libration center. This figure shows the ejection rate along a path where the initial value of $M_2$ is always equal to $316^{\circ}$ (solid line with white circles) while the "elliptic segment"  ($M_2 = 340^{\circ}$) is represented with dashed lines with black circles.   While the ejection rate along the elliptic segment is almost maximal, the solid line indicates the presence of a more stable region. Indeed, an increase of stability, denoted by a sharp change in the stability index (with a minimum of about $0.6$) arises when $a_2$ belongs to the interval $[8.28:8.33]\, \au$, which fits very well with one of the most stable regions (some kind of libration center) inside the 2:1 planetary MMR (see Fig. \ref{fig:res_plan}). It turns out that the possibility to find stable Trojans when the two planets are in MMR depends strongly on the geometry of this resonance. 

In order to demonstrate that, except the three above-mentioned peaks, the main unstable structures (ejection peaks) are generated by \FII, we will develop three different arguments. 
 First, a straightforward reasoning will show that the three other families cannot induce these instabilities. The second argument will be based on the very good agreement between the simulations presented in Fig. \ref{fig:prediction_FII}.a-c and the predictions regarding the location of the resonances of \FII.
 The last argument will consist on a detailed study of the region lying between $9.26$ and $9.4 \au$.

First, we know that \FIV acts only on a small neighborhood of a given MMR. This prevents this family from producing instabilities like the one generating the largest and highest peak of Fig. \ref{fig:prediction_FII}.c.  
Regarding \FI, its resonances depend mainly on $n_1$ which varies only  very little in our simulations. Consequently, the resonances of \FI have approximately the same strength along the migration path and thus cannot be responsible of these successive stability changes. 
The same argument holds for the secular resonances of \FIII.  The secular resonance most affecting Jovian Trojans is the $s= s_2$ resonance. It moves continuously  (except in a small neighborhood of the MMRs) without  coming close to $L_4$.  Other higher order secular resonances involving nodal secular frequencies can eventually move inside the Trojans phase space, but generate only a very local chaos (see section \ref{sec:crossingFIV52}). 
A very interesting point is connected to the existence of the secular resonances $g= (k+1)g_1 - kg_2$, which  can play an important role in increasing the eccentricity of the Trojans belonging to. On the base of these values of $g_1$ and $g_2$ given in Fig. \ref{fig:frec_plan} and according to formulas (\ref{eq:bounds}), the exact location of these resonances can be derived.  Tab. \ref{tab:res_sec} gives the locations of the main resonances of this kind. According to Fig. \ref{fig:prediction_FII} these values correspond to a region which is close to the 2:1 MMR, and  where very  high peaks are present. We come back to this point later, but in any case, the influence of secular resonances seems to be very local and cannot generate this succession of highly unstable regions. 
  
   Having excluded the significance of other families for the peaks of the ejection rate in Fig. \ref{fig:prediction_FII}, only \FII remains. This is not very surprising for at least two reasons.  The first one is that we have shown in section \ref{sec:RNp2BP} that these resonances  were able to act far away from their associated MMR. In fact, \pu reports that, in the present Solar system, narrow unstable regions resulting from \FII appear in $\Theta_{I^{\star}}$. The second reason  lies in the fact that in \cite{MoLeTsiGo2005}, the authors point out two strong instabilities which enable the capture of the future Jupiter Trojans. According to the authors, these instabilities correspond to regions where $\nu \approx 3\ineg12$ and $\nu \approx 2\ineg12$. Obviously, these resonances are members of our \FII.   
In order to make sure that these peaks are really due to \FII, we will compare, as in section \ref{sec:crossingFIV52}, our predictions to the simulations for which results are plotted in Fig. \ref{fig:prediction_FII}. 
To predict the locations of Saturn where resonances of \FII  penetrate the Trojan swarms, we proceed as for \FIV : neglecting the right-hand side of equation (\ref{eq:FIIp}), that is, the combination of secular frequencies, the resonance condition reads: 
$\nu = \frac{j}{i}\ineg{\alpha}{\beta}$. The question is now to know whether or not, for a given $a_2$, the frequency $\frac{j}{i} \ineg{\alpha}{\beta}$ belongs to $\pi_{\nu}(\Theta_{I^*})$. The bottom frame of Fig. \ref{fig:prediction_FII} answers the question: for the three main MMRs, namely $(\alpha,\beta) \in \{(1,2), (3,7), (2,5)\}$, the values $\frac{j}{i} \ineg{\alpha}{\beta}$ are plotted versus $a_2$. 
The X-axis represents $a_2$ in $\au$ while the vertical axis measures the frequencies in $\asay$.  A color is associated to each MMR, the curves $\frac{j}{i} \ineg{\alpha}{\beta}$ (which are practically straight lines) are drawn in red for the \FII connected to the 2:1 MMR, in black for the one associated to the 7:3 MMR and in green for the 5:2 MMR.  The black labels on these lines indicate the values of $\frac ij$ associated to the corresponding resonance. 
For the sake of clarity, these labels are aligned. The upper line (located at about $8600\asay$) corresponds to $\ineg12$, the second line (at $8200$) to $\ineg37$, and the last one (at $7600$) to $\ineg25$.
The colors of the background correspond to the values of the initial inclination $I^*$. Grey is associated to $I^*=2\de$: more precisely, in the grey region, the frequencies (Y-axis) range in the interval $\pi_{\nu}(\Theta_{2\de})$. In the same way, in the pink rectangle, the frequencies belong to $\pi_{\nu}(\Theta_{20\de})$, and to $\pi_{\nu}(\Theta_{30\de})$ for the cyan domain. The use of these different colors allows to predict easily the location of resonances for the three inclinations used in this paper.

As was the case for \FIV, the criterion $\frac ij\ineg{\alpha}{\beta} \in \pi_{\nu}(\Theta_{I^*})$ does not correspond to a single resonance in the Trojans phase space but rather to a multiplet of resonances associated to the linear combination of secular frequencies arising in the right hand side of (\ref{eq:FIIp}).
Consequently narrow strips should replace the lines. But for the sake of simplicity, we prefer to keep the lines.

In Fig.\ref{fig:prediction_FII}.a-c, between the 2:1 MMR (left) and $8.8 $ \au, three high peaks stand out for all inclinations (the ejection rate is practically equal to 100\%) . As mentioned above, the resonances of \FII associated to the 2:1 MMR dominate this region; therefore, these three strong depletion zones are generated by these resonances. Indeed, these peaks coincide perfectly with the lines labeled with 3, 2 and 1 (bottom frame) associated to the relations:  $\nu$ equals 3, 2 and 1 times $\ineg12$. These maxima of ejection match very well the prediction: the bases of the peaks coincide with the projection on the X-axis of the intersection of the corresponding line with the colored region (i.e. pink for $I^*= 2¡\de$). Moreover, the peaks are shifted rightwards when the inclination grows, which is due to the fact that red lines have positive slopes. 
We have to mention that it is not the first time that these structures are observed. Indeed, in \cite{MoLeTsiGo2005}, the authors have widely studied the region corresponding more or less to the interval $[8.3, 8.65] \, \au$ in our simulations (see next section). They found two regions of strong depletion corresponding to our first two peaks. Moreover, in Fig. 1 of the above-mentioned paper, the curve indicating the fraction of population that survives for $2\, 10^5$ My in the co-orbital region possesses a small singularity (just after $1.5$ My of integration) which is not mentioned in the paper. Looking at Fig. \ref{fig:prediction_FII}, we clearly see that this singularity is generated by the subfamily $\nu =  \frac32\ineg12$.

The main peak corresponding to the encounter of the secondary resonance $\nu = \ineg12$, is outside of the region studied in \cite{MoLeTsiGo2005}, but it seems that an important phenomenon occurs here. Indeed, the rightwards shift of this structure when $I^*$ increases is more pronounced than in the two previous peaks. This happens essentially because the slope of the red curve representing the frequency  $\nu = \ineg12$ is smaller than the ones of the curves $\nu = 2\ineg12$ and $\nu = 3\ineg12$. F. Marzari and H. Scholl (\citeyear{MaScho2007}) mention the existence of this resonance but observe that its influence is weaker than that of the previous secondary resonances (the 1:2 and 1:3). It is not what we get in our study, but the two simulations are not easy to compare. The first simulation takes into account a forced migration, imposing Trojans to cross a given resonance quite rapidly. While in our case, the migration being frozen, the resonances have enough time (here 10 My) to eject the test-particles.  In any case, we probably overestimate the influence of the resonances, but our goal is not to have a realistic simulation of Jupiter's Trojans during planetary migration,  but to point out the regions or the events that are relevant in term of ejection (or injection) of bodies in the co-orbital region. 
In the same paper Marzari and Scholl stress the dominant role of the secular resonance $g = g_2$.  In our simulation, the region where $g = g_2$ occurs is so close to the 2:1 MMR  (see Table \ref{tab:res_sec}) and so chaotic that  the fundamental frequencies are meaningless: Trojan trajectories are too far from quasiperiodic ones.  But, on the other hand, for $I^*=2\de$, we can distinguish a very small peak centered at $a_2 = 8.44\au$ (arrows in Fig.\ref{fig:prediction_FII}.a-c) which is generated by the secular resonance $g=3g_2-2g_1$.  By increasing $I^*$, the secular resonance is shifted rightwards (Table \ref{tab:res_sec}) while the secondary resonance moves leftwards. This generates an overlap  at approximately $20\de$ (peak in Fig.\ref{fig:prediction_FII}.b). These two regions are split again at higher inclination (Fig.\ref{fig:prediction_FII}.a).  Nevertheless the dynamical influence of the resonance $\nu = 3\ineg12$ seems to be always predominant. 

When the semi-major axis of Saturn is greater than $8.8 \au$, the dynamical situation becomes richer but more complicated. Indeed, the resonances of \FII generated by the frequency $\ineg12$ are still present, but although they are involved in the global dynamics of the present Solar system (see \pu), their dynamical influence decreases strongly.  On the contrary, the secondary resonances of \FII associated to $\ineg37$ and above all $\ineg25$ begin to generate instabilities. 
On one hand, the order of the 2:5 and 3:7 MMRs being greater than the one of the 1:2 MMR, the corresponding resonances of  \FII will probably perturb the Trojan swarms less than the ones depending on the 2:1 MMR. On the other hand, one can expect to find strong chaos when at least two of these resonances overlap. Lets us notice that, because $\ineg37 = \ineg12 +\ineg25$, when a resonance of  \FII associated to  $\ineg12$ and  another  one associated to $\ineg25$ overlap (intersection of red lines and green lines in  Fig.\ref{fig:prediction_FII}.d) a secondary resonance connected to $\ineg37$  is present in the same place.  But generally, the order of this resonance is quite high. 
 As we have retained only the resonances $\nu = \pm \ineg37$ and $2\nu =  \pm \ineg37$ (the resonances associated to $\nu = \pm 2\ineg37$ are also of interest, but for the sake of clarity they are not drawn in the figure)  only four triple intersections are inside the studied domain (Fig. \ref{fig:prediction_FII}.d).
 These intersections are defined by the relations:

\begin{figure*}
\begin{tabular}{cc}
\includegraphics[width=8.8cm,height=12cm]{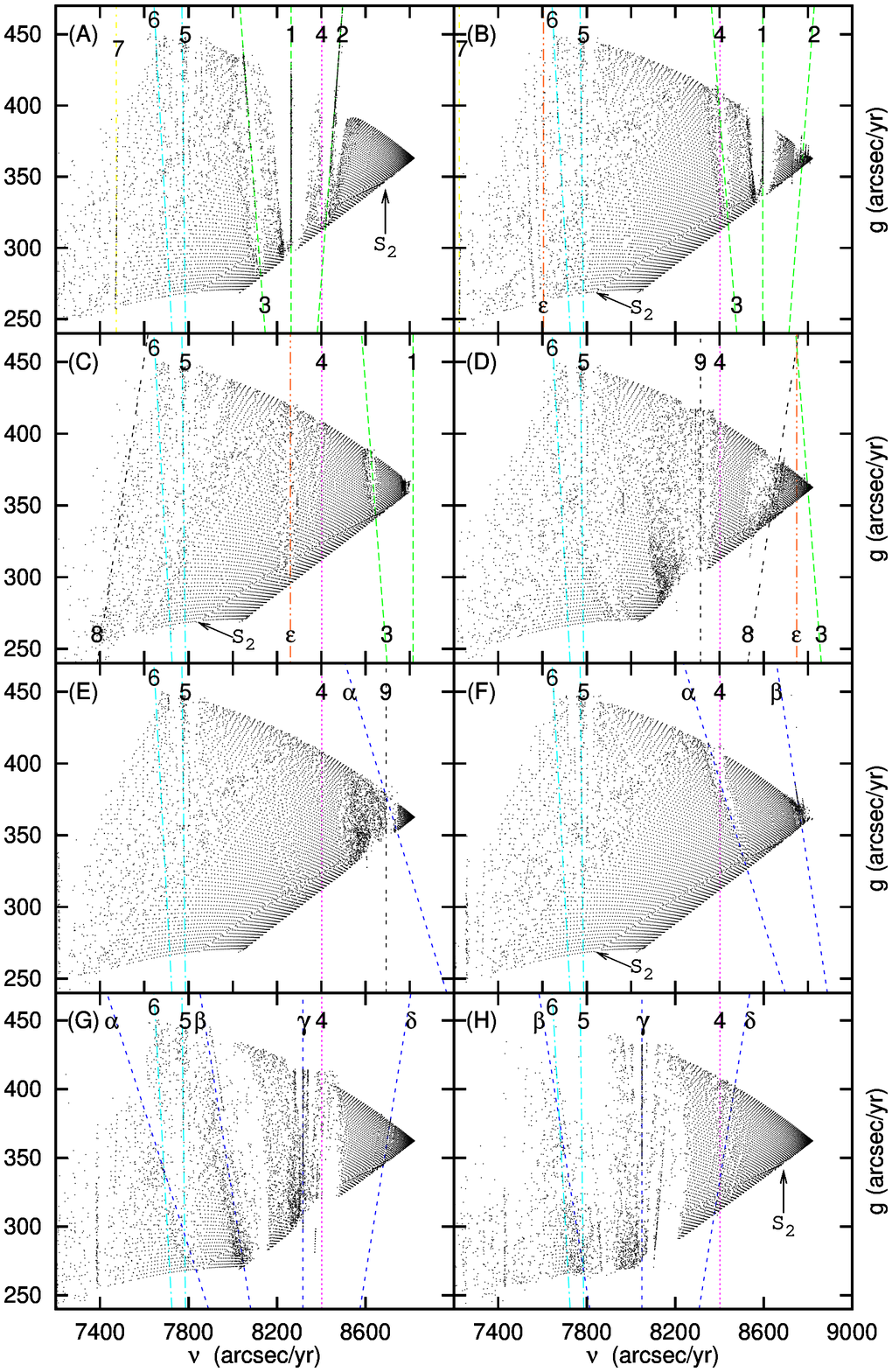} &
\includegraphics[width=8.8cm,height=12cm]{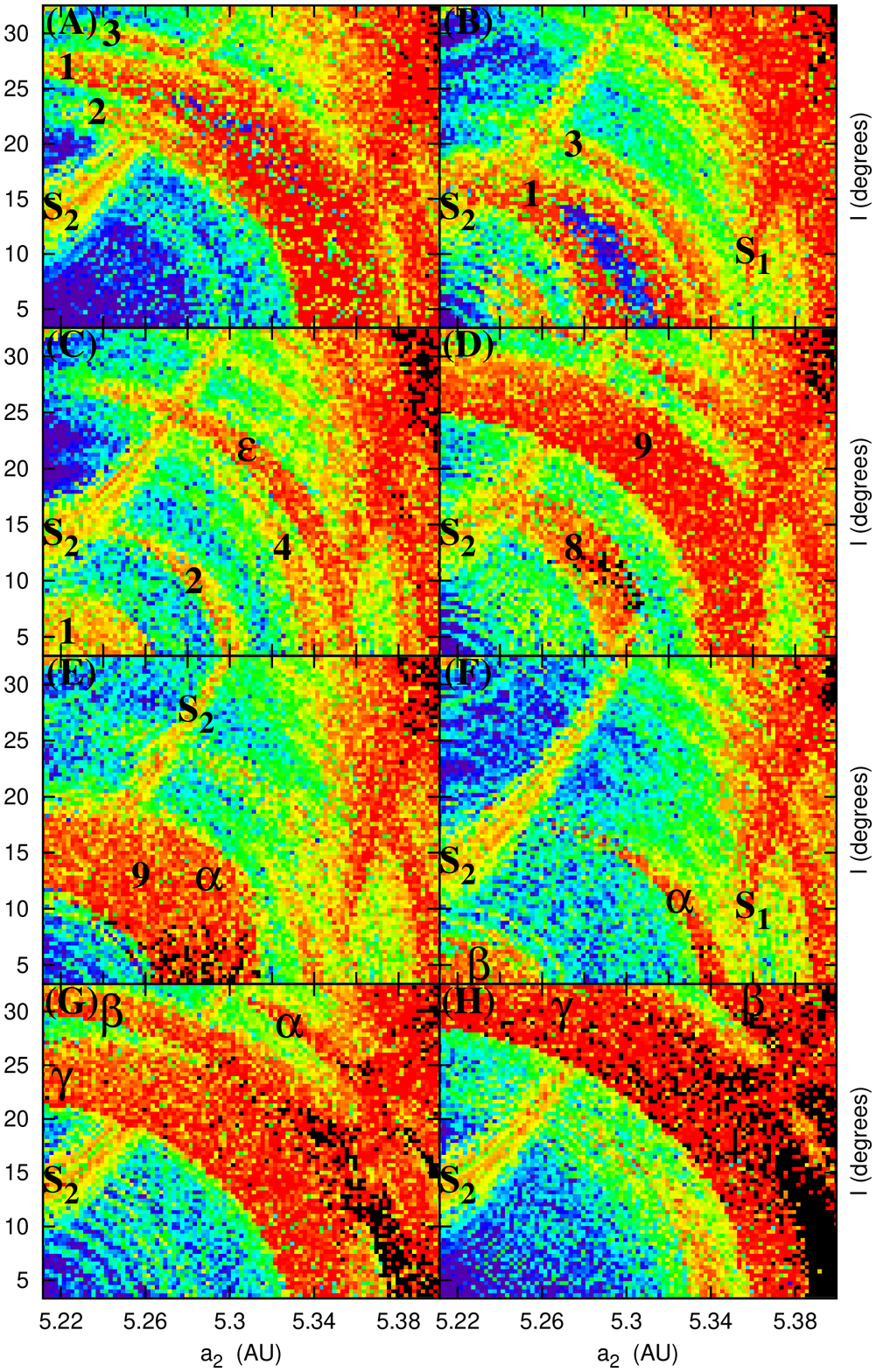}\\
\end{tabular}
\caption[]{Destabilization of the Trojan swarms by  the resonances of  \FII. Left block: Dynamical maps in the frequency space (projection on the $(\nu,s)$ plane).  Right block: Dynamical maps in the action space  (the initial conditions of the fictitious Trojans are chosen in the $(a,I)$ plane, the four other elliptic elements being fixed). The color cod associated to the diffusion rate is the same as in Fig. \ref{fig:frec_FIV_I02} .}
 \label{fig:detail_FII}
\end{figure*}

\begin{eqnarray}
\nu = \frac23 \ineg12 &= -\frac25 \ineg25   &= -\ineg37            \quad    \text{Êat about }\quad 9.04 \au \label{eq:overlap-a}\\
\nu =\frac25 \ineg12 &= - \frac23\ineg25   &= \phantom{\frac12}\ineg37               \quad    \text{Êat about }\quad 9.33 \au \label{eq:overlap-b} \\
\nu =\frac12 \ineg12 &= - \ineg25               &= \phantom{\frac12}\ineg37               \quad    \text{Êat about }\quad 9.37 \au  \label{eq:overlap-c} \\
\nu =\frac25 \ineg12 &= -2 \ineg25             &= \frac12\ineg37   \quad  \text{Êat about }\quad 9.5 \au  \label{eq:overlap-d}
\end{eqnarray}
These four different overlaps coincide with local maxima of the ejection rate. The most striking phenomenon occurs when the resonances of \FII associated to $\nu = \ineg37$ cross the Trojans' phase space (see formulas (\ref{eq:overlap-b}) and (\ref{eq:overlap-c})), for $a_2 \in [9.33,9.37] \au$. 
Indeed, at $I^* = 30\de$ (Fig.\ref{fig:prediction_FII}.a), a steep peak rises around $9.41\au $ and reaches $80\%$, while a smaller one is located on its left at $9.3\au$. When the initial inclination falls from $30\de$ to $2\de$, these two unstable structures merge to generate the rounded peak around $9.37 \au$ (Fig.\ref{fig:prediction_FII}.c). The predictions given in Fig.\ref{fig:prediction_FII}.d suggest that the left above mentioned structure is generated at $30\de$ by the triple resonance defined by equation (\ref{eq:overlap-b}), while the right one is generated  by the intersection of the $2\nu = \ineg12 $ and the $\nu = -\ineg25$ secondary resonances. When $I^*$ decreases, the higher peak follows the \FII resonances associated to $\nu = -\ineg25$ and is consequently shifted leftwards, while the other one is driven by the $\nu = \ineg12$ and goes rightwards.
The merging of those two structures occurs at $I^* = 2\de$ (Fig. \ref{fig:prediction_FII}.c) and generates the high rounded peak around $9.37 \au$ (see eaqation (\ref{eq:overlap-c})). The last resonance maintains the relatively high level of ejection on the left side of the peak.

\begin{table}
\caption{Main resonances crossing the Trojan swarms for $a_2 \in [9.26,9.4] \au$. The first column gives the label of each simulation, these labels are reported in Fig.  \ref{fig:detail_FII}. The initial values of $a_2$ adopted in each simulation are presented in the second column. The labels and the definitions of the resonances emphasize in Figs. \ref{fig:detail_FII}.A-H are given in the two last columns.  }
\begin{tabular}{c c c l }
\hline
\noalign{\smallskip}
  Figure & $ a_2 \, (\au)$  &  Label & Resonance  \\  
  A & $  9.26 $  & 1 & $\ineg12 -2\nu + g_1 = 0$  \\  
  - &  $  - $        & 2 & $\ineg12 -2\nu + g = 0$  \\  
  - &  $  - $        & 3 & $\ineg12 -2\nu - g + 2g_1 = 0$  \\  
  - &  $  - $        & 7 & $\ineg13 + 4\nu + 2s_2 = 0$  \\  
  B & $  9.30421 $  & 1, 2, 3 &  defined above \\  
  - &  $  - $        & $\eps$ & $\ineg49 -3\nu +2g_1 +3g_2 = 0$  \\  
  C & $  9.33368 $  & 1, 3, $\eps$ & defined above   \\  
  - & $  - $  & 8  & $\ineg37 -\nu +g +3g_1=0$   \\  
  D & $  9.35579 $  & 3, $\eps$, 8&  defined above \\  
   - & $  - $  & 9 & $\ineg37 -\nu +4g_2 =0$   \\  
  E & $  9.36316 $  & 9 & defined above \\ 
   - & $  - $  & $\alpha$ &   $\ineg25 +\nu + 2g + g_2 = 0$ \\  
  F & $  9.37053 $  & $\alpha$ &  defined above  \\  
    - & $  - $  & $\beta$  & $\ineg25 +\nu + g -g_1 + 3g_2 = 0$   \\  
  G & $  9.39263 $  & $\alpha, \beta$ & defined above   \\  
  - & $  - $  & $\gamma$  &  $\ineg25 +\nu  + 3g_2 = 0$  \\  
  - & $  - $  & $\delta$      &  $\ineg25 +\nu - g + 2g_1 + 2g_2 = 0$  \\  
  H & $  9.4 $  & $\beta$, $\gamma$, $\delta$ & defined above   \\  
  A to H & $9.26$ to $9.4$  & $S_1$ & $s -s_2  = 0$  \\  
      -     & $   $  & $S_2$ & $s -s_2 + g_1 - g_2 = 0$  \\  
      -     & $   $  & 4 & $  n_1 - 13\nu  - g_2 = 0$  \\  
      -     & $   $  & 5 & $ n_1 - 14\nu  - g = 0$  \\  
      -     & $   $  & 6 & $ n_1 - 14\nu  -5g + 4g_2 = 0$  \\  
\end{tabular}
\label{tab:res_FII}
\end{table}

In order to verify that the mechanism that has just been presented is the most relevant one and that it is really \FII which drives the global dynamics of this region, we will present a sequence of simulations in the same spirit as in section \ref{sec:crossingFIV52}.  Indeed, if the resonances of \FII play an important dynamical role, they should be visible in the frequency space, and particularly on the $(\nu,g)$ projection. Although we have seen in sections \ref{sec:model} and \ref{sec:crossingFIV52} (see also \pu) that $\pi_{(g,s)}\circ {\cal F}$ (resp. $\pi_{(\nu,s)}\circ {\cal F}$) which is the composition of the frequency map with the projection on the $(g,s)$ plane (resp. on the $(\nu,s)$ plane) defines a isomorphism from the regular regions of ${\cal D}_{I^*}$ to $\pi_{(g,s)}(\Theta_{I^*})$ (resp. $\pi_{(\nu,s)}(\Theta_{I^*}$)),  the projection $\pi_{(\nu,g)}\circ {\cal F}$ on the $(\nu,g)$ plane is not one to one \citep{GaJoLo2005}.  Consequently a point on the $(\nu,g)$ plane possesses several pre-images making the identification of resonances very difficult.  
To overcome this difficulty, we have decided to cut the phase space by the plan of coordinates $(a, I)$, limited to the rectangle ${\cal D'} = [5.21,5.4]\times[3^{\circ},33^{\circ}]$, where the other initial conditions satisfy $(e^{(0)}, \lambda^{(0)}, \varpi^{(0)},\Omega^{(0)}) = (e^{(0)}_1, \lambda^{(0)}_1 + \pi/3, \varpi^{(0)}_1 +\pi/3,\Omega^{(0)}_1)$. Denoting $\Theta'$ the image of ${\cal D'}$ by the frequency map, the map $\pi_{(\nu,g)}\circ {\cal F}$ now defines  a isomorphism between ${\cal D'}$ and $\pi_{(\nu,g)}(\Theta')$. This is clearly shown by Fig. \ref{fig:detail_FII}.
As Fig. \ref{fig:frec_FIV_I02}, Fig. \ref{fig:detail_FII}  is composed of two blocks. The right block corresponds to dynamical maps of the domain ${\cal D'}$ for eight different values of the initial semi-major axis of Saturn. These eight values are given in Table \ref{tab:res_FII}. The color code associated to the diffusion rate being the same as in Fig. \ref{fig:frec_FIV_I02}, it is not reported here. 
The left block is the corresponding view in the frequency space. Here again, the correspondence between the $(a,I)$ plane and $(\nu,g)$ plane is easy to establish. We start from $L_4$ (bottom left corner in the $(a,I)$ plane) which corresponds to the right vertex of $\pi_{(\nu,g)}(\Theta')$  where $(\nu ,g) = (8820,362)$. When $a_2$ increases, $I$ remaining constant, $\nu$ decreases  while $g$ increases up to the point $(\nu ,g) = (7680, 451)$. On the contrary, always starting at $L_4$, if $I$ increases when $a$ is constant, both $\nu$ and $g$ decrease down to the point of coordinates $(8050, 270)$. The frequency map being continuous, it is easy to deduce where the images of the two other edges of the rectangle ${\cal D'}$ are located.     
In the frequency space (left block) vertical, or nearly vertical,  structures are dominant. As we will see later, these structures are associated to resonances of \FII of the form $\nu + k g +c =0$, where the absolute value of the integer $k$ is lower than $2$ and $c$ is a constant real number. On the contrary, it is important to notice that no horizontal line is visible. This confirms the fact that \FIV does not have significant influence on the dynamics of this region. 
Although there exists a priori no direct relation between the ejection rates presented in Fig. \ref{fig:prediction_FII} and the present simulations, for the reason that the first ones are deduced from integrations in the $(a,e)$ plane while the second ones concern the $(a,I)$ plane, these new simulation illustrate very well the predictions presented in Fig. \ref{fig:prediction_FII}.d. But before comparing these predictions to the numerical simulations, let us first recall that the location of some families on either the $(a,e)$ or $(a,I)$ planes is  practically independent of the value of the initial semi major axis of Saturn $a_2$. This is the case for the secular resonances of \FIII (at least in the studied interval $[9.26,9.4] \au$) and especially for \FI. Consequently these resonances are present at the same place along the sequence. 
But the intersections of the resonances of \FIII with the $(a,I)$ plane are different from their intersections with  the $(a,e)$ plane. Consequently their shape will be different here than in section \ref{fig:frec_FIV_I02}. While their intersections are similar to arches (arcs of ellipses centered in $L_4$) in the $(a,e)$ plane, in the $(a,I)$ plane their shapes are comparable to the curves plotted in Fig. \ref{fig:schema}. In particular, the resonance $s - s_2 = 0$ is quite similar to the curve in  Fig. \ref{fig:schema} labeled with $1$, while the resonance $s -s_2 + g_1 - g_2 = 0$ is well represented by the curve $3$. In the right block of Fig. \ref{fig:detail_FII}, the resonance $s -s_2 + g_1 - g_2 = 0$, denoted $S_2$ in Fig. \ref{fig:detail_FII} and Table \ref{tab:res_FII} , is always clearly visible above $I = 10^{\circ}$  while the resonance $s - s_2 = 0$ (denoted $S_1$) stands out only in Fig. \ref{fig:detail_FII}. H, E and F. In the other plots of the sequence,  $S_1$ and other resonances of \FII merge in the red region lying in the right part of the figures. 
On the other hand, the identification of these two resonances is not so easy in the frequency projection on the $(\nu,g)$ plane. The presence of $s -s_2 + g_1 - g_2 = 0$ can be deduced from the existence of the curve, located at the bottom of the frequency domain,  which corresponds to singularities of the frequency map.  This curve is labeled with $S_2$. 
The resonance $s-s_2=0$ is only detectable by the instability that it generate in the $(\nu,g)$ plane. But as these resonances do not play a significant role in the mechanism we want to describe,   it is not necessary to give more details. For the same reason, we will not discuss any longer the shape and the location of this structure. For the sake of completeness, three resonances of \FI are represented in the frequency space , labeled respectively with $4,5$ and $6$ , for the combinations  $  n_1 - 13\nu  - g_2 = 0$, $ n_1 - 14\nu  - g = 0$ and $ n_1 - 14\nu  -5g + 4g_2 = 0$. In the $(a,I)$ plane, the resonances $4$ and $5$ are not discernible for they are located close to $5.4\au$ in the same chaotic region than the resonance $S_2$, while the resonance $4$ is located around $a_2 = 5.34\au$ (see Fig .\ref{fig:detail_FII}.C right block).

According to Fig. \ref{fig:prediction_FII}.d, if we start  the migration from $a_2 = 9.26\au$ and increase the initial semi-major axis of Saturn up to $9.4\au$, the first noticeable resonances of \FII encountered are the ones defined by the relation $\ineg12 -2\nu + \cdots = 0$, where the dots represent combinations of secular frequencies. 
Indeed, in Fig. \ref{fig:detail_FII}.A left, three resonant lines stand out very clearly. They are emphasized by red lines labeled with $1, 2$ and $3$ (the associated combination of frequencies is given in Table \ref{tab:res_FII}). When $a_2$ slightly increases, these resonances move towards higher values of $\nu$, which is in perfect accordance with our predictions. The resonances related to $2\nu = \ineg12$ leave permanently the Trojan swarms in Fig. \ref{fig:detail_FII}.D for $a_2 = 9.35579 \au$. 
In the $(a,I)$ space, these resonances and more generally the resonances of \FII appear as elliptically shaped structures centered on $L_4$. In the simulations A to C, these  structures move towards $L_4$ before leaving the phase space.  It is worth mentioning that these instabilities are not generated by a single resonance but  by a multiplet of resonances which overlap. This is particularly visible in the frequency space, where parallel resonant lines are observable in chaotic regions  (this phenomenon is detailed in \pu). But, for the sake of clarity, only one resonant line is drawn in these structures. 
According to Fig.  \ref{fig:prediction_FII}.d, when the resonances considered above leave the phase space (red  line labeled with 1/2), the subfamily associated to $\nu = \ineg37$ begins to enter the swarm. This is exactly what happens in simulation C where the black line (denoted 8) corresponding to the resonance $\ineg37 -\nu +g +3g_1=0$ appears on the left hand side of the $(\nu,g)$ plane (see Fig. \ref{fig:detail_FII}). 
 As for the resonances $2\nu = \ineg12$ which cross the phase space towards $L_4$, the subfamily associated to $\ineg37$ enters the swarm  in Fig. \ref{fig:detail_FII}.C (label $8$) and is still present in simulation E (label $9$). Among the eight simulations that we performed, the simulation E, which corresponds to $a_2 = 9.36316 \au$, is perhaps the most interesting. Indeed, just before the $\ineg37$ subfamily exits through the $L_4$ point, the resonances denoted $\alpha, \beta, \gamma$ and $\delta$ enter the swarm. These new resonances corresponding to the relation $\nu = -\ineg25$ move in the opposite direction than the previous ones. This corresponds to the fact that, in Fig. \ref{fig:prediction_FII}.d, the green curve (labeled with $-1$) associated to $\nu = -\ineg12$ is decreasing while the previous one (the black curve labeled with $1$) is increasing. 
 It turns out that these two structures collide giving rise to the broad red arch surrounding $L_4$ (label $9$ and $\alpha$ in Fig. \ref{fig:detail_FII}.E). The three last simulations show the evolution of the resonances of this last subfamily when $a_2$ increases up to $9.4 \au$, which validates one more time the predictions given in section \ref{sec:FII}.

The study of this region proves definitely that the resonances of \FII dominate its dynamics, and that, as it was mentioned in the beginning of this section, this family generates the main part of the peaks of instability present in Fig. \ref{fig:prediction_FII}a-c.
Before completing this section, let us mention that two other resonances, which are rather anecdotic, appear in the previous simulations. Resonance  $7$, which is defined by $\ineg13 + 4\nu + 2s_2 = 0$ is observable in simulations A and B. In fact,  in \pu   we have already encountered resonances  of \FII associated to the frequency $\ineg13$ at high inclination. On the other hand, it is much more surprising to realize that the resonance $\ineg37 -\nu +g +3g_1=0$ denoted $\epsilon$ stands out against the background of the frequency space and generates its own unstable structure visible on the $(a,I)$ plane at least in Fig.\ref{fig:detail_FII}.C.

\subsection{Dependence on the initial phases}
\label{sec:phases}
 \begin{figure}
\includegraphics[width=8.5cm,angle=0]{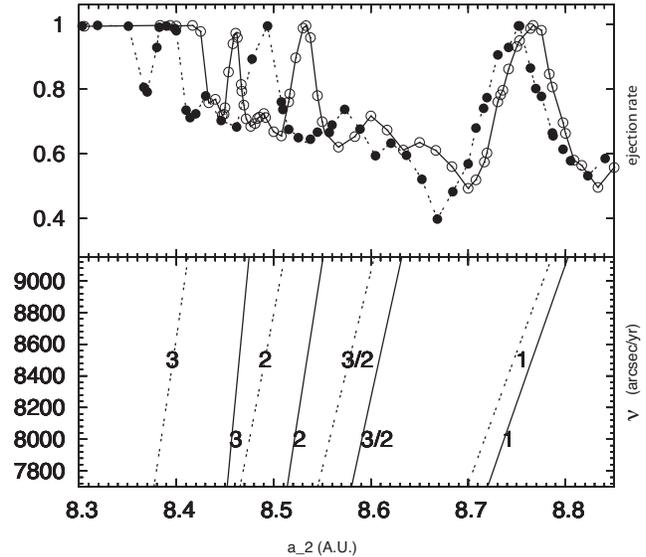}
\caption[]{ Dependence of the ejection rate on the initial phases for the resonances of \FII: the bottom frame shows the predicted locations of the resonances $\nu \approx 3\ineg12$,  $\nu \approx 2\ineg12$, $\nu \approx 3/2\ineg12$ and $\nu \approx \ineg12$ (see Fig. \ref{fig:prediction_FII}-d), for the "elliptic section" in solid lines and for the "hyperbolic section" in dashed lines. The top-frame shows the ejection rate for these two sections.}
\label{fig:prediction_FII_hyp}
\end{figure}

 Adiabatic invariance theory suggests that, during migration, the planets  are never captured in MMRs \citep{MoLeTsiGo2005}.
 For this reason, and even if we do not take migration into account, we will study the evolution of the resonant structure along the "hyperbolic segment" defined in section \ref{sec:plan-freq}. Contrarily to the "elliptic segment", the former has the advantage to cross the 2:1 resonance at its smallest section, minimizing the range of Saturn's semi-major axis for which the two planets are trapped inside the MMR. This simulation is probably closer to what is expected by adding planetary migration.
 
Denoting by $\ineg12^H$ the quantity $n_1 - 2n_2$ evaluated along the hyperbolic segment (Fig.\ref{fig:frec_plan} green curve) and $\ineg12^E$ the one computed on the elliptic path (Fig.\ref{fig:frec_plan} red curve), we deduce  from  Fig.\ref{fig:frec_plan} that when $a_2\in [8.3, 8.85]$ (outside of the MMR), the inequality $\ineg12^E < \ineg12^H$ holds. 
 Moreover, the greater the value of $a_2$, the smaller the value of $\ineg12^H - \ineg12^E$.  In addition, the width (in $a_2$) of the 2:1 MMR along the two sections are very different: the "elliptic width" is about three times greater than the hyperbolic one (see also Fig. \ref{fig:res_plan}). 
 As a result, the slope of the red curve is always greater than the slope of the green one. 
 These facts have direct consequences on the secondary resonances of \FII associated to $\ineg12$. 
 Fig.\ref{fig:prediction_FII_hyp} illustrates the comparison between these two choices of initial phases. In its bottom frame, the frequencies $3, 2, 3/2$ and $1$ times $\ineg12$ are represented by solid lines for  $\ineg12^E$ and by dashed lines for $\ineg12^H$. Just like Fig\ref{fig:frec_plan}-d, this plot enables us to predict the locations of the resonances of \FII. The associated numerical simulations are reported in the top frame of  Fig.\ref{fig:prediction_FII_hyp} \footnote{In this section, the simulations are limited to the Trojans having an initial inclination equal to  $2^\circ$, since one example is sufficient for the purpose of studying the dependence of the results on the planets' initial phase.}, the ejection rates along the hyperbolic section are represented by a dashed curve with black circles while the solid line with white circles corresponds to the elliptic section. Again, the numerical simulations and the predictions are in very good agreement.  
 As regards the location of the resonances of \FII, a given resonance is crossed by the Trojan swarm at a smaller value of $a_2$ in the hyperbolic case than in the elliptic one, and the splitting between these two locations decreases when the planetary system moves away from the orbital resonance.
The width of the resonances of \FII is also affected by the change in the initial phases in the planets. Because the slope of the curves plotted in the bottom frame of Fig. \ref{fig:prediction_FII_hyp} is larger for the elliptic section than for the hyperbolic section, the resonances generated in the hyperbolic case are wider (i.e. they occur within larger intervals $\Delta a$) than the ones associated to the elliptic case.  
But, as shown in Fig. \ref{fig:prediction_FII_hyp}, even if the ejection peaks are wider and shifted leftwards along the hyperbolic section, the global features are the same in both cases. 
Consequently, the modification of the planetary phases does not introduce appreciable effects on  the global dynamics of the Trojan swarms. 

\section{Discussion}
\label{sec:Discussion}

The global dynamics of a Trojan swarm is shaped by its resonant structure. This structure depends crucially on the fundamental frequencies of the planetary system in which it is embedded. 
  By generalizing the results of \pu concerning Jupiter's Trojans, we show that in a general planetary configuration this resonant structure can be decomposed in four different families.  
  Among these families, two are particularly sensitive  to the geometry of the planetary system: \FII and \FIV are dependent on the closeness of the planetary system to MMRs. It turns out that variations of the semi-major axis of the planets deeply modify the dynamical influence of these two families. 
 
  Based on this decomposition and on the dependence of the families on fundamental frequencies of the planetary system, we present a general method making possible the study of the global stability of Trojan swarms during planetary migration possible. We show that the knowledge of the evolution of the fundamental frequencies of the system during its migration suffices  to predict the main features of its global dynamical evolution (transition between total stability and strong instability), and to find  which planetary configurations which lead to stable or unstable swarms.

In our application to the case of the Jupiter Trojans disturbed by Saturn, we have varied one parameter: the semi-major axis of Saturn\footnote{ Equivalent results are obtained if one varies, instead, both semi-major axes, i.e. the results depend only on the ratio of the two axes.}.
While in a realistic migration, many other parameters would vary, this would modified only slightly the results obtained in this paper.
  For example, the variations of the planetary phases have almost no effect on our results (section \ref{sec:phases}).  In the same way, the "apsidal corotation" of the planetary perihelia which, according to \cite{MaScho2007}, strengthens the influence of the secular resonances, probably does not significantly modify the resonant structure of the Trojans. 
In section \ref{sec:crossingFIV52} we have described the mechanism leading \FIV to generate strong instabilities when the planetary system approaches a MMR.

Section \ref{sec:FII} analyzes the dominant role of \FII in the production of unstable regions during planetary migration. In particular, we demonstrate a mechanism of formation of huge  chaotic regions by the merging of several resonances crossing the Trojan swarm in opposite directions. 
 
The resonant structure, and particularly its splitting in four families as defined in section \ref{sec:RNp2BP} is generic. Moreover, the definitions of the families seem to hold for almost all planetary configurations. But the respective dynamical influences of the four families depend on the mass of the planet harboring the Trojan swarms. According to formula (\ref{eq:freq_L4}), if this mass decreases, the decrease in the proper frequency $g$ is proportional to the planetary mass while the one in $\nu$  is weaker (proportional to the square root of the planetary mass). 
 The elements of \FIV being approximated by the relation: $g \approx \ineg{p}{q}$, the smaller the planetary mass, the closer this resonance is to the $p:q$ MMR.  As a result, the smaller the mass of the planet, the more local the dynamical  influence of \FIV.  In this case, we can expect that the role of the resonances of \FIV becomes negligible compared with the role of \FII. Conversely, the secular resonances (\FIII) which do not involve the frequency $s$, seem to be negligible in the Jovian Trojan swarm. But their contributions increase for smaller planetary masses. This has been already mentioned in section \ref{sec:RNp2BP}.
This has been observed by  \citet{Bo2008}  (available at http://www.imcce.fr/page.php?nav=en/publications/theses/\, index.php) in the case of the Trojans of Saturn. In both cases, \FII and \FIV dominate the dynamics of the swarms during migration.  

Finally, the method of prediction developed in this paper being very general, it can be used to study the dynamics of a large class of problems. Beside its straightforward application to the Trojans of a planet (results regarding the dynamics of the hypothetical Trojans of Saturn will be presented in a forthcoming paper), it can also be used to study the dynamics of satellites in a planetary system in migration, or more generally, to understand the behavior of a dynamical system undergoing quasiperiodic perturbations with slowly varying frequencies.

\subsection*{Acknowledgments}
We are deeply indebted to the anonymous referee for his precious suggestions regarding the structure of this paper and for stressing some inconsistencies in earlier versions. 
We wish to acknowledge Gwena'l Bou\'e and especially Fran\c cois Farago for critical reading of draft versions.  
This work has been partially supported by PNP-CNRS. The computing clusters IBM-SP4 at CINES  have been widely used.


\newcommand{\noopsort}[1]{}




\end{document}